\documentclass[aps,prd,twocolumn,superscriptaddress,longbibliography,amsmath,amssymb,amsfonts,citeautoscript]{revtex4-1}
\usepackage{amsmath}
\usepackage{amssymb}
\usepackage{bm}
\usepackage{bbm}
\usepackage{color}
\usepackage{epstopdf}
\usepackage{graphicx}
\usepackage[urlcolor=blue,colorlinks=true,citecolor=blue,linkcolor=blue,pdfstartview={FitH},bookmarks=false]{hyperref}
\usepackage{todonotes}
\usepackage{ulem}

\begin{document}
\title{Quench dynamics of correlated quantum dot proximitized to superconducting lead}

\author{K. Wrze\'sniewski}
\email[e-mail:]{wrzesniewski@amu.edu.pl}
\affiliation{Faculty of Physics, A. Mickiewicz University,
             61-614 Pozna\'n, Poland}

\author{B. Baran}
\affiliation{Institute of Physics, M. Curie-Sk\l{}odowska University,
             20-031 Lublin, Poland}

\author{R. Taranko}
\affiliation{Institute of Physics, M. Curie-Sk\l{}odowska University,
             20-031 Lublin, Poland}

\author{T. Doma\'nski}
\email[e-mail:]{doman@kft.umcs.lublin.pl}
\affiliation{Institute of Physics, M. Curie-Sk\l{}odowska University,
             20-031 Lublin, Poland}

\author{I. Weymann}
\email[e-mail:]{weymann@amu.edu.pl}
\affiliation{Faculty of Physics, A. Mickiewicz University,
             61-614 Pozna\'n, Poland}

\date{\today}

\begin{abstract}
Quantum system abruptly driven from its stationary phase can reveal nontrivial dynamics upon approaching a new final state. We investigate here such dynamics for a correlated quantum dot sandwiched between the metallic and superconducting leads, considering two types of quenches feasible experimentally. In particular, we examine an interplay between the proximity induced electron pairing with correlations caused by the on-dot Coulomb repulsion. We discuss the time-dependent charge occupancy, complex order parameter, transient currents, and analyze evolution of the subgap quasiparticles which could be empirically observed in the tunneling conductance.
\end{abstract}

\maketitle

\section{Motivation}
Upon bringing a quantum impurity/dot close to a bulk superconductor the quasiparticle bound states can develop inside the pairing gap $\omega \in \left( -\Delta, \Delta\right)$ \cite{balatsky.vekhter.06}. These in-gap states originate either (1) from the proximity effect, when the Cooper pairs penetrate such nanoscopic object converting it into superconducting grain, or (2) by pairing the quantum dot electron with the opposite spin electron of a bulk superconductor. Depending on the specific mechanism, they are dubbed the Andreev \cite{Rodero-11} or Yu-Shiba-Rusinov bound states \cite{Paaske-2010}, respectively. The subgap quasiparticles have been observed in numerous experimental studies, using magnetic impurities deposited on superconducting substrates \cite{STM-1,STM-2,STM-3,Franke-2018} and quantum dots embedded into the Josephson \cite{Josephson-1,Josephson-2,Josephson-3},  Andreev \cite{Andreev-1,Andreev-2,Andreev-3} or multi-terminal heterojunctions \cite{multiterminal-1,multiterminal-2,Baumgartner-17}.

With the advent of time-resolved techniques such bound states could be nowadays studied, inspecting their dynamical properties. Some aspects concerning this issue have been so far investigated theoretically by several groups, e.g.\  addressing  the response time to a step-like pulse \cite{Xing-07}, the time-dependent multiple Andreev (particle-to-hole) reflections \cite{Stefanucci-10}, sequential tunneling \cite{Konig-12}, influence of time-dependent bias \cite{Pototzky-14}, waiting time distributions manifested in the nonequilibrium transport \cite{Governale-13,Michalek-17}, short-time counting statistics \cite{Konig-16}, realization of the metastable bound states in the phase-biased Josephson junction \cite{LevyYeyati-2016,LevyYeyati-2017}, transient effects caused by forming the Andreev \cite{Taranko-2018} and Josephson \cite{Taranko-2019} junctions, bound states of the periodically driven systems \cite{Komnik-13,Melin-2017,Arachea-2018,Baran-2019}, cross-correlations between currents of a Cooper pair splitter \cite{Wrzesniewski-2020,Flindt-2020,Michalek-2020} and studying more exotic heterostructures, hosting the Majorana modes \cite{Souto-2020,Jonckheere-2020,Manousakis-2020}.

Time-dependent change of the model parameters is usually followed by a thermalization processes \cite{Polkovnikov-2011,Freericks-2014}. In the superconducting heterostructures the rate of relaxation processes depends on a continuum \cite{LevyYeyati-2017}. In particular, when the quantum system is {\it quenched} from its ground state, i.e. when some parameter of the Hamiltonian is suddenly changed, the resulting time evolution might lead to nontrivial behaviour upon reaching its new asymptotic state,  sometimes undergoing the dynamical quantum phase transitions \cite{Heyl-2018}. Dynamics triggered by such {\it quantum quench}, when the initially prepared state $\left| \Psi(t_{0})\right>$ described by the Hamiltonian $\hat{H}_{0}$ undergoes evolution to $\left| \Psi(t)\right> = e^{-i\hat{H}t/\hbar} \left| \Psi(t_{0})\right>$, where at later times $t>t_{0}$ the Hamiltonian $\hat{H}\neq \hat{H}_{0}$, has been recently the topic of intensive studies. Such phenomena can be conveniently explored in nanoscopic heterostructures, because the available experimental methods enable controllable change of the system's parameters  $\hat{H}_{0} \rightarrow \hat{H}$.

\begin{figure}
\includegraphics[width=0.7\columnwidth]{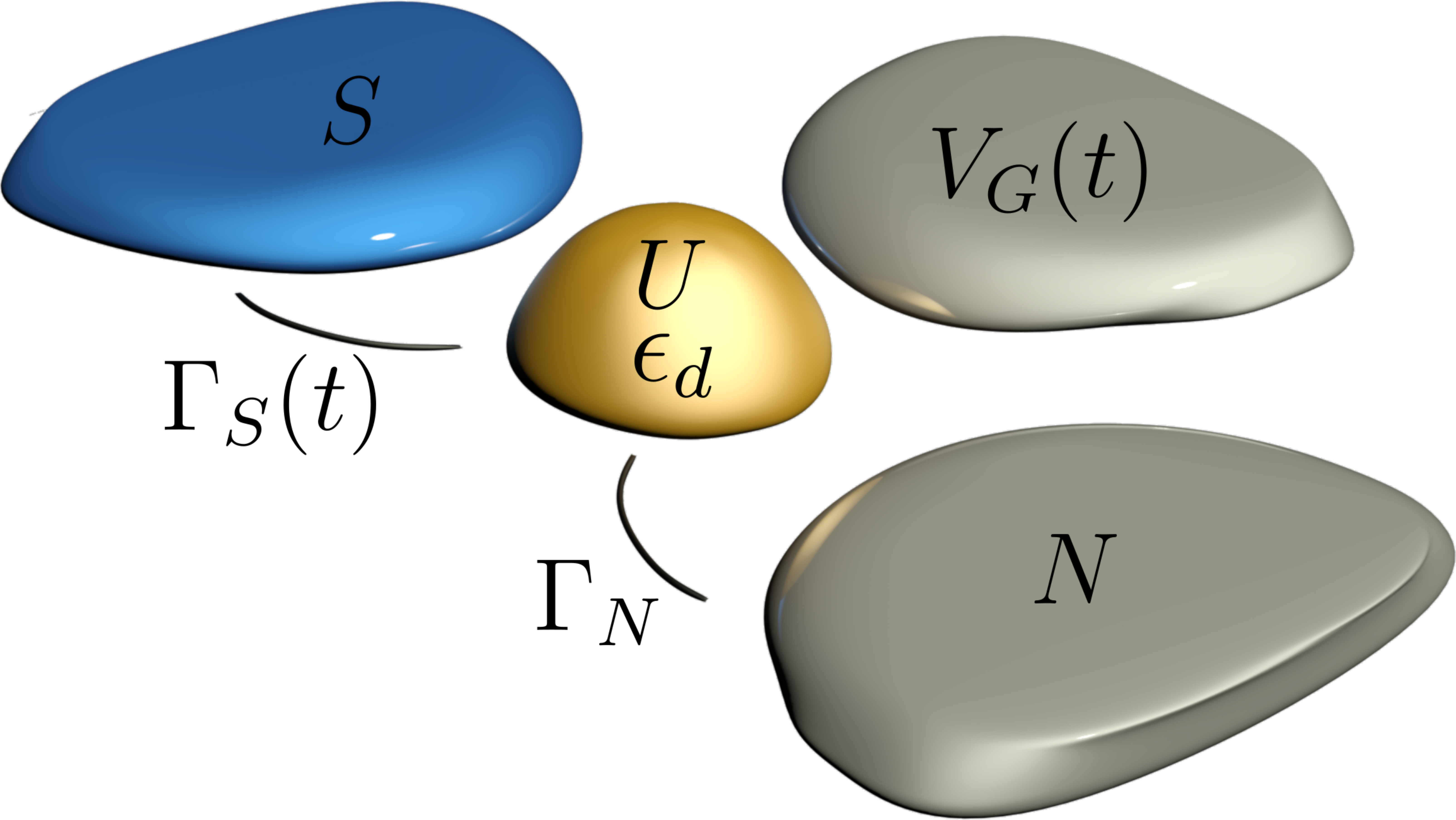}
\caption{Heterostructure, consisting of a correlated quantum dot (QD)
coupled to the normal (N) and superconducting (S) leads whose energy level $\varepsilon_d$ can be changed by the gate potential $V_{G}(t)$.}
\label{scheme}
\end{figure}

In this work we study the dynamical features of a correlated quantum dot (QD) placed between the normal (N) and superconducting (S) electrodes (Fig.~\ref{scheme}), focusing on two types of quenches caused by: (i) an abrupt change of the coupling to the superconducting lead and (ii) a sudden alternation of the gate potential lifting the energy level of QD. This allows us to explore the dynamical properties of the subgap  quasiparticles existing in a fully correlated quantum dot junction, and to examine their behavior in a vicinity of the singlet-doublet quantum phase transition.
We achieve this goal by employing the time-dependent numerical renormalization group (tNRG) method \cite{Wilson1975,Anders2005,Anders2006}
to quantitatively study the quench dynamics of an unbiased junction. On the other hand in the case of a biased heterostructure, the dynamics is examined by determining the equation of motion of relevant operators within the mean-field approximation, upon establishing the validity of such approach by comparison with tNRG in the relevant transport regime.
This enables us to draw conclusions about the dynamical behavior
of superconductor-proximized, correlated quantum dot junction subject to arbitrary bias voltage.

The paper is organized as follows. In Sec.~\ref{sec:formulation} we formulate the microscopic model, describe the specific quench protocols, and outline two computational methods for determination of the time-dependent physical observables. Next, in Sec.~\ref{unbiased_junction}, we analyze evolution of the quantum dot occupancy, the complex order parameter and the charge current induced by both quantum quenches in the unbiased heterojunction. Sec.~\ref{Sec:biased} presents the charge transport properties for the biased system, which could be suitable for experimental verification. Finally, in Sec.~\ref{Sec:conclusions}, we summarize the main results.

\section{Formulation of the problem
\label{sec:formulation}}

In this section we present the microscopic model and specify
two types of quantum quenches that could be practically realized.
Next, we outline the computational methods suitable to account for
the time-dependent phenomena, proximity effect and electron correlations.

\subsection{Microscopic model
\label{sec:micro-model}}

For the description of our N-QD-S heterostructure we use the single
impurity Anderson Hamiltonian
\begin{eqnarray}
\hat{H} =  \underbrace{\sum_{\sigma} \varepsilon_{d}(t) \hat{d}^{\dagger}_{\sigma}
\hat{d}_{\sigma}  +  U \; \hat{n}_{\uparrow} \hat{n}_{\downarrow}}_{\hat{H}_{QD}}  +
\sum_{\beta} \left( \hat{H}_{\beta} + \hat{V}_{\beta - QD} \right)
\label{model}
\end{eqnarray}
where $\hat{d}_{\sigma}$ ($\hat{d}^{\dagger}_{\sigma}$) is the annihilation (creation)
operator of the quantum dot electron with spin $\sigma$ whose (time-dependent)
energy is $\varepsilon_{d}(t)$ and $U$ denotes electrostatic repulsion between the
opposite spin electrons. We treat the external metallic lead as free fermion gas
$\hat{H}_{N} \!=\! \sum_{{\bf k},\sigma} \xi_{\bf k} \hat{c}_{{\bf k} \sigma}^{\dagger}
\hat{c}_{{\bf k} \sigma}$, where $\xi_{\bf k}=\varepsilon_{\bf k}-\mu_{N}$ is the
energy $\varepsilon_{\bf k}$ of itinerant electrons measured from the chemical potential
$\mu_{N}$. The superconducting lead is described by the BCS model $\hat{H}_{S} \!=\!
\sum_{{\bf q},\sigma}  \xi_{{\bf q}} \hat{c}_{{\bf q}\sigma}^{\dagger}  \hat{c}_{{\bf q}
\sigma} \!-\! \sum_{\bf q} \Delta  \left( \hat{c}_{{\bf q} \uparrow} ^{\dagger}
\hat{c}_{-{\bf q} \downarrow}^{\dagger} + \hat{c} _{-{\bf q} \downarrow} \hat{c}_{{\bf q}
\uparrow }\right)$ with $\xi_{\bf q}=\varepsilon_{\bf q}-\mu_{S}$ and the isotropic
pairing gap $\Delta$.

Coupling of the QD electrons to the metallic lead is given by the hybridization term
$\hat{V}_{N-QD} = \sum_{{\bf k},\sigma} \left( V_{\bf k} \; \hat{d}_{\sigma}^{\dagger}
\hat{c}_{{\bf k} \sigma} + \mbox{\rm h.c.} \right)$ and  $\hat{V}_{S - QD}$ can be
expressed by interchanging the indices ${\bf k} \leftrightarrow {\bf q}$. In the present
study we focus on the subgap quasiparticle states, therefore for simplicity we impose
the constant auxiliary couplings $\Gamma_{N (S)}=\pi \sum_{{\bf k}({\bf q})}
|V_{{\bf k}({\bf q})}|^2 \;\delta(\omega \!-\! \varepsilon_{{\bf k}({\bf q})})$.
For the energy regime $|\omega|\ll\Delta$ the coupling $\Gamma_{S}$ can be regarded
as the proximity induced pairing potential, whereas $\Gamma_{N}$ controls the inverse
life-time of the in-gap quasiparticles. As we shall see, these couplings
manifest themselves in the dynamical quantities in qualitatively different ways.

\subsection{Quench protocols}
\label{sec:quench}

Any type of the quantum quench can be generally cast
into the following time-dependent Hamiltonian
\begin{equation}
  \hat{H}(t) = \theta(-t)\hat{H}_{0} + \theta(t)\hat{H},
\label{Eq:Hamiltonian_TD}
\end{equation}
where $\theta(t)$ is the step function. The initial Hamiltonian $\hat{H}_0$ is suddenly replaced (at time $t=0$) by the new Hamiltonian $\hat{H}$. In particular, an abrupt change can be realized within the same structure of the model (\ref{model}) by appropriately modifying its parameters.

Evolution for the time-dependent expectation value of the physical observable ${\hat{\cal{O}}}(t)$ is then governed by (for time-independent Hamiltonian)
\begin{eqnarray}\label{Eq:O}
O(t) \equiv \langle {\hat{\cal{O}}}(t) \rangle = \mathrm{Tr}\left\{e^{-i\hat{H}t} \hat{\rho}_0 e^{i\hat{H}t} {\hat{\cal{O}}} \right\} \nonumber \\
=\mathrm{Tr}\left\{\hat{\rho}_0 \hat{\cal{O}}_{H}(t)\right\}\equiv\langle \hat{\cal{O}}_{H}(t) \rangle,
\end{eqnarray}
where $\hat{\rho}_0$ denotes the initial equilibrium density matrix
of the system described by $\hat{H}_0$ and $\hat{\cal{O}}_{H}(t)$ is the Heisenberg representation of $\hat{\cal{O}}$.
In this work we shall examine the dynamical behavior of various  quantities,
considering two different types of the quantum quenches. In the first case, we
impose an abrupt change of coupling to the superconducting lead
\begin{eqnarray}
V_{{\bf q}}(t) =
\left\{ \begin{array}{ll}
0 & \hspace{0.5cm} \mbox{\rm for } t \leq 0  \\
V_{{\bf q}}  & \hspace{0.5cm} \mbox{\rm for } t > 0
\end{array} \right.
\label{abrupt_coupling}
\end{eqnarray}
which is formally equivalent to the assumption $\Gamma_{S}(t)=\Gamma_{S} \; \theta(t)$.
Another type of the quantum quench will refer to the time-dependent QD  energy level
\begin{eqnarray}
\varepsilon_{d}(t) =
\left\{ \begin{array}{ll}
\varepsilon_{d} & \hspace{0.5cm} \mbox{\rm for } t \leq 0  \\
\varepsilon_{d} + V_{G}  & \hspace{0.5cm} \mbox{\rm for } t > 0
\end{array} \right.
\label{abrupt_gate}
\end{eqnarray}
which could be practically achieved  by applying the gate potential
$V_{G}(t)=V_{G} \; \theta(t)$.
For computing the time-dependent expectation values of our interest, such as the charge occupancy
$n_{\sigma}(t)\equiv \langle \hat{d}^{\dagger}_{\sigma}(t)\hat{d}_{\sigma}(t) \rangle$,
the complex order parameter $\chi(t)\equiv \langle \hat{d}_{\downarrow}(t)\hat{d}_{\uparrow}(t) \rangle$
and the charge currents $j_{S,N}(t)$ we use two techniques. Below we briefly outline both methods.

\subsection{Mean field approach}

In absence of correlations ($U\!=\!0$) one could exactly determine all required
observables, solving the set of coupled equations of motion for appropriately chosen operators. But even for $U\!=\!0$, the observables have far from trivial evolution. For the abrupt coupling of the uncorrelated QD to both external electrodes we have recently inspected the characteristic time-scales manifested in a buildup
of the subgap bound states  \cite{Taranko-2018,Taranko-2019}. Technically,
we have solved the Heisenberg equation of motion
for the localized $\hat{d}_{\sigma}^{(\dagger)}$
and itinerant $\hat{c}_{{\bf k}/{\bf q}\sigma}^{(\dagger)}$ electron operators,
respectively. For this purpose we have expressed the Heisenberg equations of motion
introducing the Laplace transforms $\hat{O}(s) = \int_{0}^{\infty} e^{-st}\hat{O}(t)dt $,
%
%
which are suitable for considering the specific initial conditions
$\hat{O}(0)$. Next, performing the inverse Laplace transforms we  have determined the time-dependent operators $\hat{O}(t)$ and used
them for computing  analytical exact formulas for the expectation values, such as $n_{\sigma}(t)
\equiv \langle \hat{d}_{\sigma}^{\dagger}(t)\hat{d}_{\sigma}(t)\rangle$.

\begin{figure}
\includegraphics[width=0.5\columnwidth]{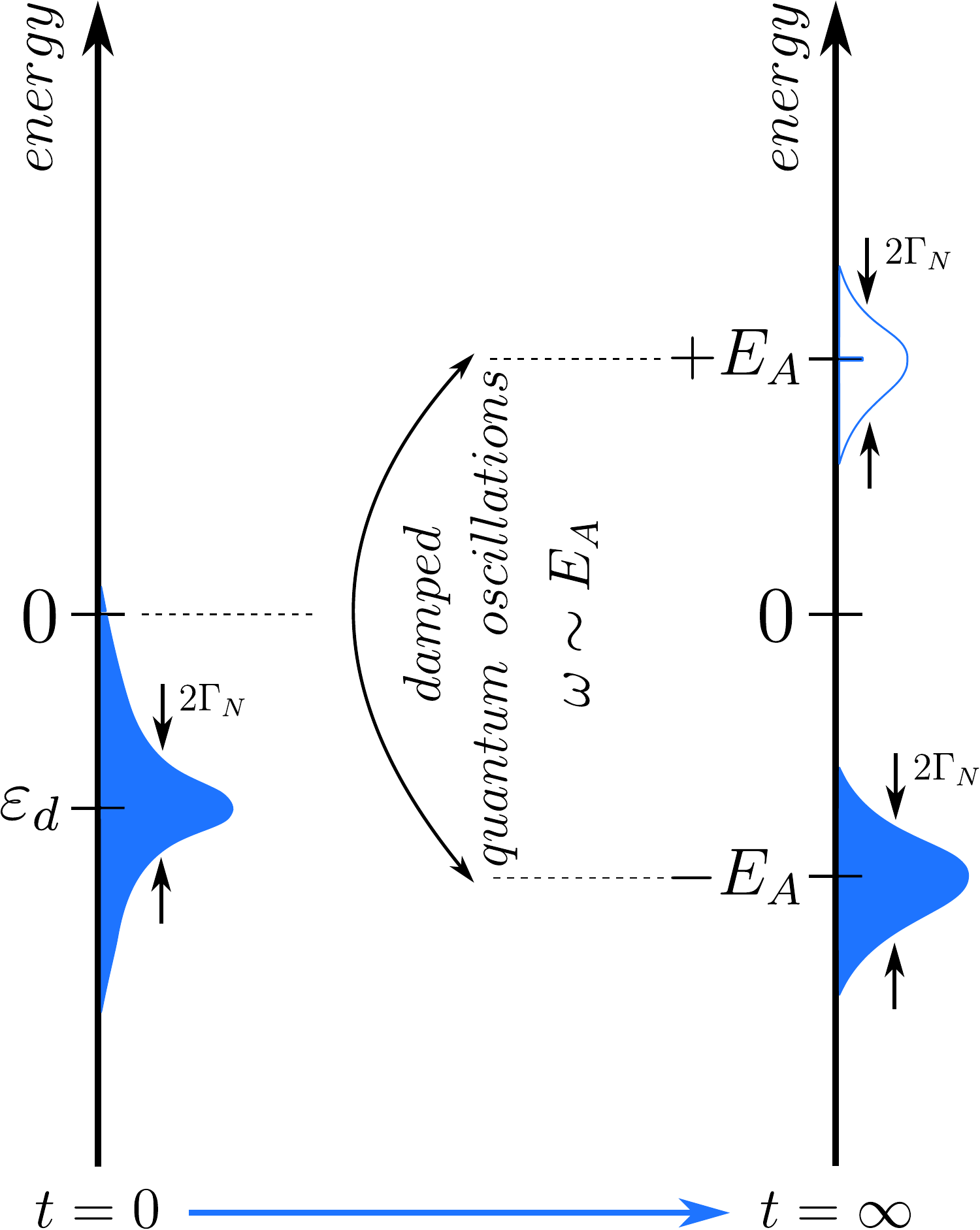}
\caption{Illustration of the post-quench evolution driven by a sudden change of the coupling to superconductor
$\Gamma_{S}(t)=\Gamma_{S}\theta(t)$, presenting the quasiparticle peak existing
till $t=0$ at $\varepsilon_{d}$, which changes into a pair of bound states at $\pm E_{A}$. Such changeover is accompanied with damped quantum oscillations of frequency $\omega=E_{A}$.}
\label{idea}
\end{figure}

Typical evolution of the uncorrelated quasiparticle spectrum driven by a sudden change of coupling $\Gamma_{S}(t)$  is schematically illustrated in Fig.~\ref{idea}. Initially, the electron state exists at the QD level $\varepsilon_{d}$ and its line-broadening (inverse life-time) depends on the coupling $\Gamma_{N}$ to the metallic bath. Upon coupling the QD to the bulk superconductor this quasiparticle state evolves into a pair of the Andreev peaks centered at $\pm E_{A}$, which for $U=0$ and $\Delta\rightarrow\infty$ are given by $E_{A}=\sqrt{\varepsilon_{d}^{2}+\Gamma_{S}^{2}}$. This new quasiparticle spectrum is gradually developed through a sequence of quantum oscillations with the characteristic frequency $\omega=E_{A}$, reminiscent of the Rabi-type oscillations of two-level systems \cite{Taranko-2018}. The relaxation processes originating from the QD coupling to the normal lead are responsible for damping of these quantum oscillations. The evolution of the time-dependent observables is thus controlled by two characteristic time scales: (i) period of the quantum oscillations $T=2\pi/E_{A}$ (ii) governed by an exponential decay $\exp{\left(-t/\tau\right)}$ with $\tau=\hbar/\Gamma_{N}$.

Similar approach leading to the analytical expressions for physical quantities does not hold for the system with the Coulomb correlations included, as the corresponding set of equations of motion for $\hat{d}_{\sigma}(t)$ and $\hat{c}_{k/q \sigma}(t)$ can not be closed. Even for a weakly correlated system, when the Coulomb repulsion term can be linearized within the Hartree-Fock-Bogoliubov (mean field) decoupling scheme
%
%
\begin{eqnarray}
\hat{d}^{\dagger}_{\uparrow}\hat{d}_{\uparrow}
\hat{d}^{\dagger}_{\downarrow}\hat{d}_{\downarrow} & \simeq &
n_{\uparrow}(t) \hat{d}^{\dagger}_{\downarrow}\hat{d}_{\downarrow}
+n_{\downarrow}(t) \hat{d}^{\dagger}_{\uparrow}\hat{d}_{\uparrow}
-n_{\uparrow}(t)n_{\downarrow}(t)
\nonumber \\ & + &
\chi(t) \hat{d}^{\dagger}_{\uparrow} \hat{d}^{\dagger}_{\downarrow}
+\chi^{*}(t) \hat{d}_{\downarrow} \hat{d}_{\uparrow}
-\left| \chi(t) \right|^{2} ,
\label{HFB}
\end{eqnarray}
the analytical approach (described above) fails. With the approximation (\ref{HFB}),
we can incorporate the Hartree-Fock term into the renormalized QD energy level
$\tilde{\varepsilon}_{d}(t) \equiv \varepsilon_{d}
(t)+U n_{-\sigma}(t)$, whereas the anomalous contribution rescales the effective
pairing potential $\tilde{\Gamma}_{S}(t) \equiv \Gamma_{S}(t) - U \chi(t)$.
In comparison to the case with $U=0$, now the effective QD level and effective pairing
potential are time-dependent functions. The corresponding equations of motion for $\hat{d}_{\sigma}(t)$ and $\hat{c}_{{\bf k}/{\bf q}\sigma}(t)$ cannot be transformed in a tractable way through the Laplace transformation into an algebraic system of equations for $\hat{d}_{\sigma}(s)$ and $\hat{c}_{{\bf k}/{\bf q}\sigma}(s)$ and next, through the inverse Laplace transformation into final required expressions for $\hat{d}_{\sigma}(t)$ and $\hat{c}_{{\bf k}/{\bf q}\sigma}(t)$. However, in such a case we can find the observables of interest, $n_{\sigma}(t)$ and $\langle \hat{d}_{\downarrow}(t)\hat{d}_{\uparrow}(t)\rangle$, solving numerically the set of coupled equations of motion for
$n_{\sigma}(t)$, $\langle \hat{d}_{\downarrow}(t)\hat{d}_{\uparrow}(t)\rangle$, $\langle \hat{d}^{\dagger}_{\sigma}(t)\hat{c}_{{\bf k}\sigma}(0)\rangle$ and $\langle \hat{d}_{\sigma}(t)\hat{c}_{{\bf k}-\sigma}(t)\rangle$, respectively (see Ref.~\cite{Taranko-2018}). In this paper we are going to consider different types of quantum quenches applied in the system under consideration using this method of calculations.

Obviously, one may ask about the validity of the static mean field approximation
(\ref{HFB}). This decoupling could be expected to give credible results, whenever
the Coulomb potential $U$ is much smaller than the pairing strength $\Gamma_{S}$
(recall, that on-dot pairing is driven here by the superconducting proximity effect).
Nonetheless, it has been shown \cite{Zonda-2015} that (under the stationary conditions) the lowest order treatment  (\ref{HFB}) of the Coulomb interaction
qualitatively reproduces the even-odd parity change of the ground state realized
at $U\sim\Gamma_{S}$. Such results well agree with the numerical renormalization
group data and with the quantum Monte Carlo simulations \cite{Novotny-2019}.
Motivated by this fact, in the following we confront this approximation with the sophisticated
(and computationally more demanding) time-dependent numerical renormalization group
method. While the latter method allows for accurate studies of dynamics in the strongly-correlated regime, the Hartree-Fock scheme enables the  determination of the differential
tunneling conductance in the biased heterojunction (Sec.~\ref{Sec:biased}).

\subsection{Time-dependent numerical \\renormalization group}

The time-dependent numerical renormalization group (tNRG) is an extension of the Wilson's numerical renormalization group (NRG) method,
which allows one to conveniently study the dynamics of quantum impurity systems \cite{Wilson1975,Bulla2008,Anders2005, Anders2006,Costi2014,Costi2014generalization}.
An invaluable advantage of this approach is the very accurate treatment of many-body correlations in a fully non-perturbative manner.

In order to study the quench dynamics of the system described by the time-dependent Hamiltonian specified in Eq.~(\ref{Eq:Hamiltonian_TD}),
we use the NRG method to solve both the initial and final Hamiltonians, $\hat{H}_0$ and $\hat{H}$, independently \cite{NRG_code}. In the NRG procedure both Hamiltonians are diagonalized in an iterative manner,
keeping at each iteration at least $N_K$ energetically lowest-lying eigenstates labeled with superscript $K$. The high-energy discarded states, labeled with superscript $D$, are collected from all the iterations and used to construct the full many-body
initial and final eigenbases \cite{Anders2005}
\begin{equation} \label{Eq:completeness}
\sum_{nse}|nse\rangle^{\!D}_{0} \,{}^{D}_{\,0}\!\langle nse| \!=\! \mathbbm{\hat{1}} \;\;\;\,
 {\rm and}
 \,\;\;\; \sum_{nse}|nse\rangle^{\!D} \,{}^D \!\langle nse| \!=\! \mathbbm{\hat{1}},
\end{equation}
corresponding to $\hat{H}_0$ and $\hat{H}$, respectively. The index $s$ labels the eigenstates at iteration with integer number $n$, while $e$ indicates the environmental subspace representing the rest of the Wilson chain. Here, we note that all eigenstates of the last iteration are
considered as discarded.
In the next step, an initial full density matrix $\hat{\rho}_0$ is constructed for the system
described by $\hat{H}_0$ at thermal equilibrium \cite{Andreas_broadening2007}
\begin{equation}
\hat{\rho}_0=\sum_{nse}\frac{e^{-\beta E_{0ns}^D}}{Z} |nse\rangle^{\!D}_{0} \,{}^{D}_{\,0}\!\langle nse|,
\end{equation}
where $\beta \equiv 1/T$ is the inverse temperature and
\begin{equation}
Z\equiv\sum_{nse} e^{-\beta E_{0ns}^D}
\end{equation}
is the partition function.

The actual time-dependent calculations are performed in the frequency space.
The expectation value of the frequency-dependent local operator
$O(\omega)\equiv\langle \mathcal{\hat{O}}(\omega) \rangle$
expressed with the use of the corresponding eigenstates is given by
\begin{eqnarray}\label{Eq:Ow}
    O(\omega) &=&\!\!
       \sum_{n}^{ XX'\neq KK}\sum_{n'} \sum_{ss'e}  {}^{X}\! \langle nse|w_{n'} \hat{\rho}_{0n'}| ns'e\rangle^{\! X'} \nonumber\\
   &&\times {}^{ X'}\! \langle ns'e|\mathcal{\hat{O}}|nse\rangle^{\! X} \; \delta(\omega + E_{ns}^{X} - E_{ns'}^{X'}),
\end{eqnarray}
where $\hat{\rho}_{0n'}$ is the part of the initial density matrix
given at iteration $n'$ and $w_{n'}$ is the weight of the contribution evaluated by tracing out the environmental states \cite{Andreas_broadening2007}.
The calculation of the expectation value is performed
in an iterative fashion by adding all the contributions,
as described in Ref.~\cite{WrzesniewskiWeymann-2019}.
Subsequently, the discrete data is weakly smoothed with a log-Gaussian function and broadening parameter $b \leq 0.1$, and then Fourier-transformed into the time domain to eventually obtain a time-dependent expectation value of the local operator \cite{Andreas2012}
\begin{equation}
    O(t)=\int_{-\infty }^{\infty}O(\omega)e^{-i \omega t} d \omega.
\end{equation}
For results calculated with the tNRG procedure
we used the discretization parameter $1.5 \leqslant \Lambda \leqslant 2$, set the length of the Wilson chain to $N=100$
and kept at least $N_K=2000$ eigenstates at each iteration.
More detailed description of the tNRG implementation
in the matrix product state framework has been presented in Ref.~\cite{WrzesniewskiWeymann-2019}.

\section{Dynamics of unbiased setup}
\label{unbiased_junction}

\begin{figure}
\includegraphics[width=0.95\columnwidth]{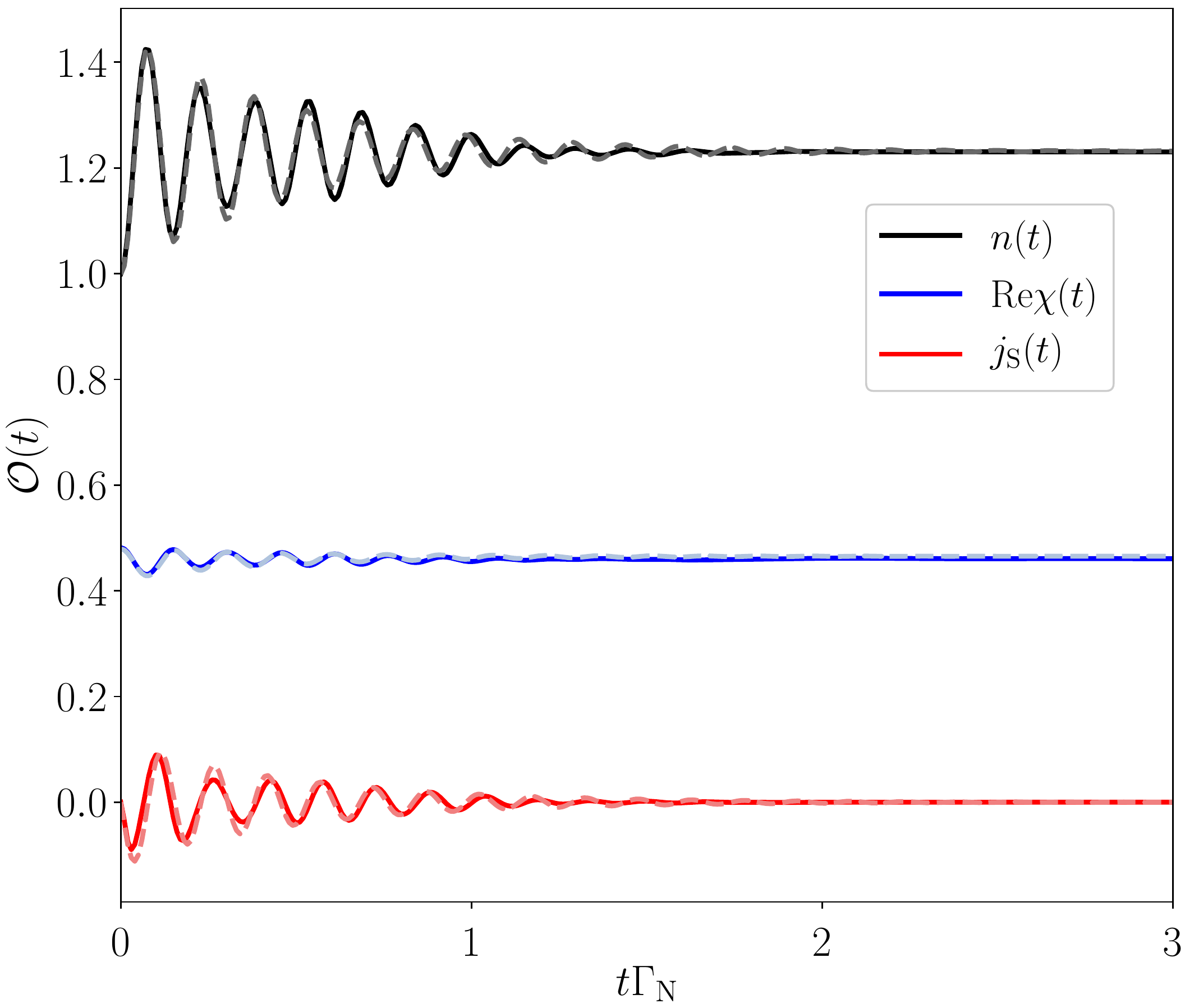}
\caption{Comparison of the time dependent observables obtained by  tNRG technique (solid lines) and HFB approximation (dashed lines) for a sudden change of the QD level from $\varepsilon_{d}(t\leq 0)=-U/2$ to $\varepsilon_{d}(t>0)=-U$. The couplings of QD to external leads are assumed to be $\Gamma_{S}=0.2$, $\Gamma_{N}=0.01$ and $U=0.1$. tNRG parameters are in units of band halfwidth.}
\label{quench_eps}
\end{figure}

\begin{figure}[t]
\includegraphics[width=0.95\columnwidth]{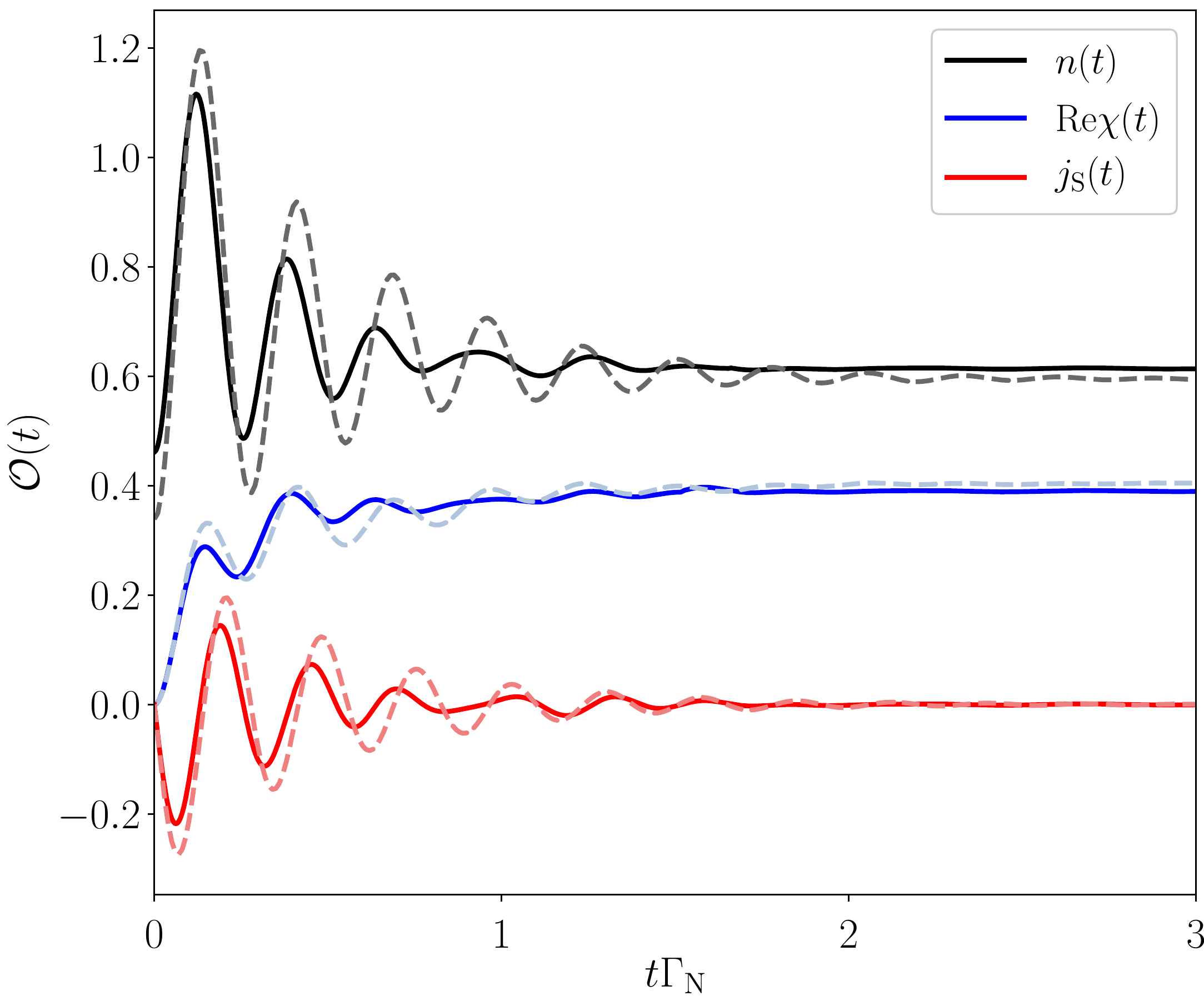}
\caption{Comparison of the tNRG results (solid lines) with the mean field values (dashed lines) obtained for $\varepsilon_{d}=0$, $U=0.1$ and $\Gamma_{N}=0.01$ imposing the quench of $\Gamma_{S}$, $\Gamma_{S}(t<0)=0\rightarrow \Gamma_{S}(t>0)=0.1$.}
\label{quench_gammas}
\end{figure}

We have checked that in the weak correlation limit, $U\ll\Gamma_{S}$, both computational procedures yield practically identical results. In what follows, we shall inspect the time-dependent quantities obtained under arbitrary conditions as a test for credibility of the approximate treatment, which will be used in Sec.~\ref{Sec:biased} to compute the transport properties of the biased heterostructure. For this purpose, we restrict our considerations to the superconducting atomic limit $\Delta \rightarrow \infty$ and assume a small coupling $\Gamma_{N}=0.01$ in order to guarantee the long life-times of in-gap quasiparticles. The latter assumption is also useful for the analysis of the relaxation processes, whose characteristic time-scale is $\tau \sim \Gamma_{N}^{-1}$ \cite{Taranko-2018}.

Figure~\ref{quench_eps} shows the time-dependent occupancy $n(t)$, charge current $j_{S}(t)$ (expressed in units of $\frac{4e}{\hbar}\Gamma_{S}$) and the real part of the order parameter $\chi(t)$ obtained for a sudden change of the QD level $\varepsilon_{d}(t)$. Since the current to the normal contact, $j_{N}(t)$, obeys the charge conservation, $j_{S}(t)+j_{N}(t)=e\frac{d n(t)}{dt}$, we skip its presentation here.  Figure~\ref{quench_gammas} displays the same quantities obtained for a sudden switching of the coupling $\Gamma_{S}(t)=U\;\theta(t)$. In both cases we clearly recognize that the initial observables gradually evolve to their new steady-state-limit values over the characteristic time interval $\tau \sim 1/\Gamma_{N}$. Meanwhile, they undergo the quantum oscillations, whose frequency depends on the energies of in-gap quasiparticles. Such behavior has been previously obtained by us analytically \cite{Taranko-2018} for the noninteracting case (see Fig.~\ref{idea}). In what follows we shall analyze the role of electron correlations.

\subsection{Quench in coupling $\Gamma_S$}

For understanding the dynamics of the correlated quantum dot driven by any type of the quench, it is useful to recall the stationary solution in the limit of $\Gamma_{N}=0$ and $\Delta \rightarrow \infty$. Depending on the model parameters, i.e. $\varepsilon_{d}$, $U$ and $\Gamma_{S}$, the quantum dot can be either in the singly occupied $\left| \sigma \right>$ or the BCS-type $u \left| 0\right> - v \left| \uparrow \downarrow \right>$ ground state  \cite{Bauer-2007}. For
\begin{eqnarray}
\left( \varepsilon_{d} + \frac{U}{2} \right)^{\!2} + \Gamma_{S}^{2} = \left( \frac{U}{2}\right)^{\!2}
\end{eqnarray}
there occurs a {\it quantum phase transition} from the (spinful) doublet to the (spinless) singlet configuration. It has crucial importance for an interplay between the on-dot pairing and the correlation effects. For finite $\Gamma_{N}\neq 0$, such transition is replaced by a crossover. Nonetheless, all the essential features of these qualitatively different (singlet/doublet) phases are still clearly observable.

\begin{figure}
\includegraphics[width=1\columnwidth]{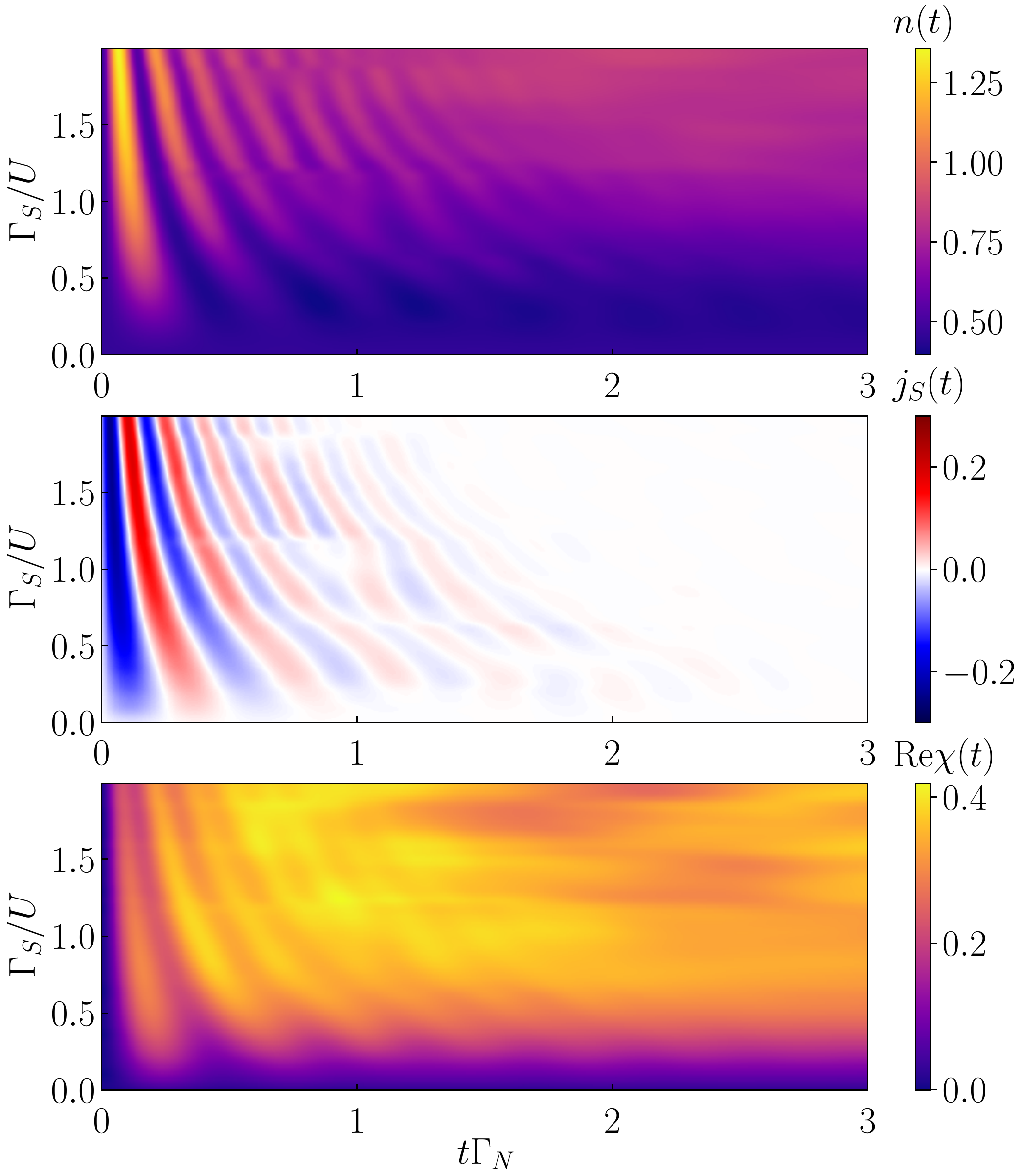}
\caption{The time-dependent occupation number $n(t)$, current $j_S(t)$  [in units $\frac{4e}{\hbar}\Gamma_{S}$] and the real part of $\chi(t) = \langle d_\downarrow(t) d_\uparrow(t) \rangle$ after switching the coupling strength $\Gamma_S(t)$
from zero to its final value $\Gamma_S$. Results are obtained by tNRG for parameters as in Fig.~\ref{quench_gammas}}.
\label{tnrg_gs_quench1}
\end{figure}
\begin{figure}
\includegraphics[width=1\columnwidth]{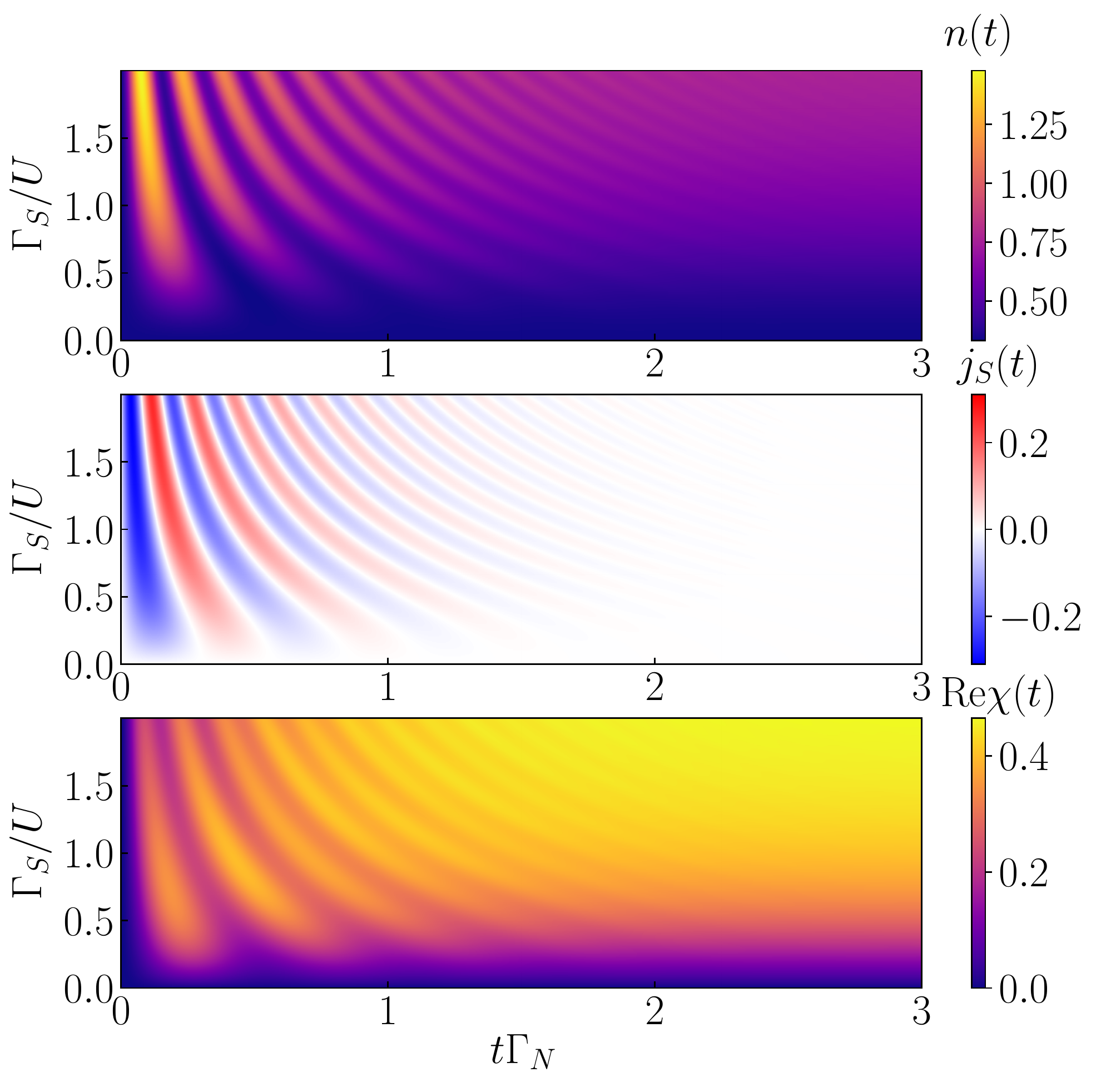}
\caption{The same as in Fig.~\ref{tnrg_gs_quench1} obtained by the mean-field approximation.}
\label{hfb_gs_quench1}
\end{figure}

\begin{figure}
\includegraphics[width=1\columnwidth]{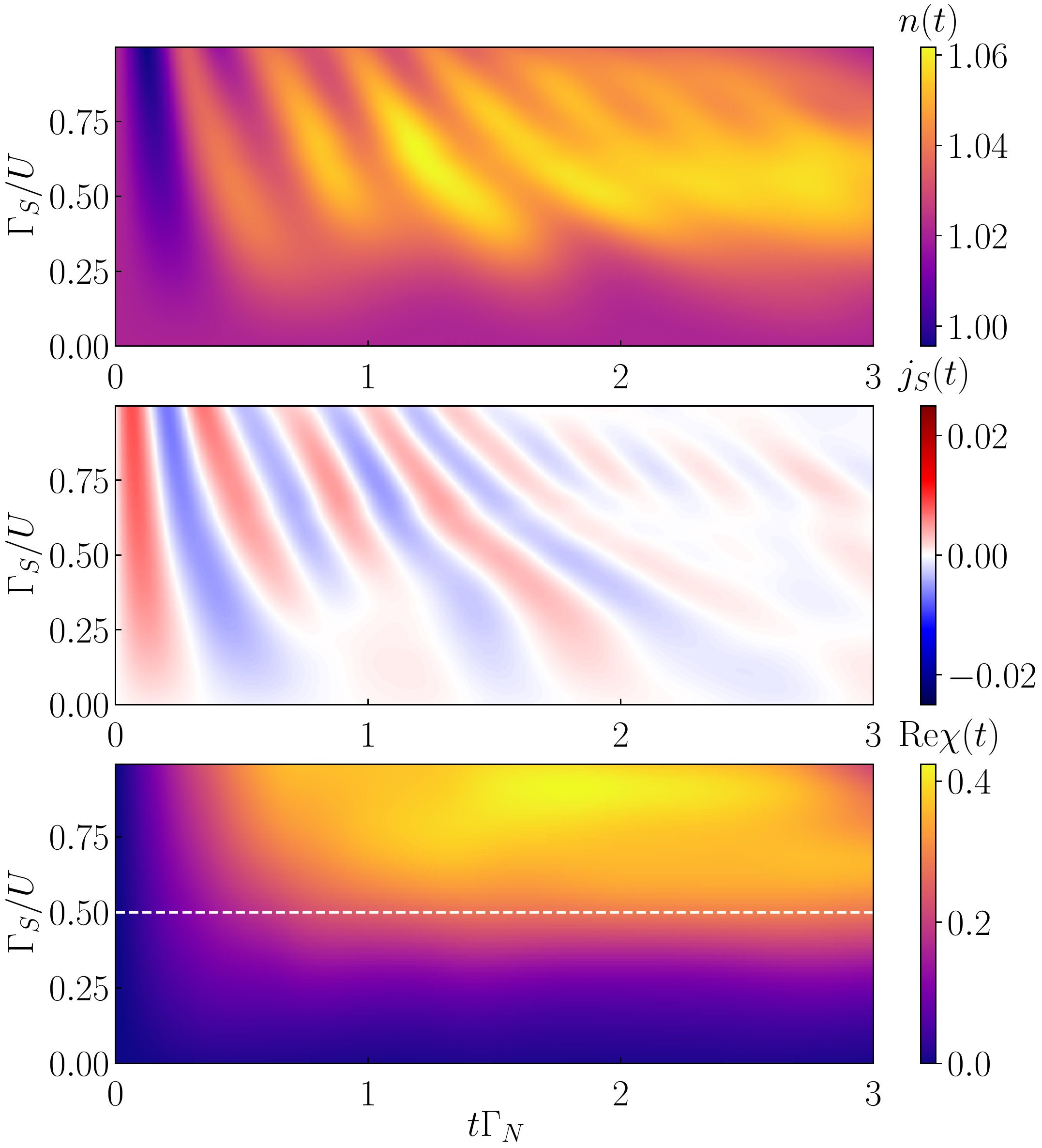}
\caption{The time-dependent occupation number $n(t)$, current $j_S(t)$ and real part of the order parameter
$\chi(t) = \langle d_\downarrow(t) d_\uparrow(t) \rangle$ after switching the coupling strength $\Gamma_S(t)$
from zero to its final value $\Gamma_S$ (indicated on vertical axis). Results are obtained by tNRG for parameters as in Fig.~\ref{quench_gammas} and $\varepsilon_{d}=-U/2 - \delta$, where $\delta=U/20$.}
\label{tnrg_gs_quench3}
\end{figure}

In particular, for the half-filled dot ($\varepsilon_{d}=-U/2$), such quantum phase transition (crossover) would occur at $\Gamma_{S}=U/2$. Figures~\ref{tnrg_gs_quench1} and \ref{hfb_gs_quench1} present the evolution of the physical quantities with respect to time (horizontal axis) and the final coupling strength $\Gamma_{S}$ (vertical axis) obtained for $\varepsilon_{d}=0$ by tNRG  and mean field approximation, respectively. In this case the quantum dot evolves to the BCS-type configuration for all values of $\Gamma_{S}$. Figures~\ref{tnrg_gs_quench3} and \ref{hfb_gs_quench3} correspond
to the nearly half-filled quantum dot. As expected, in the doublet region ($\Gamma_{S} < U/2$)  we notice the order parameter to be negligibly small (bottom panel) and we hardly observe any significant charge flow $j_{S}(t)$ (middle panel) due to the dominant Coulomb repulsion. For stronger couplings $\Gamma_{S} > U/2$, the system again evolves to the BCS-type ground state, and this is achieved through a sequence of the damped quantum oscillations. With increasing $\Gamma_{S}$, the quasiparticle energies move further and further away, therefore the oscillation frequency grows.
Let us notice, that both methods yield practically identical results.

\subsection{Quench in orbital level position}

\begin{figure}
\includegraphics[width=1\columnwidth]{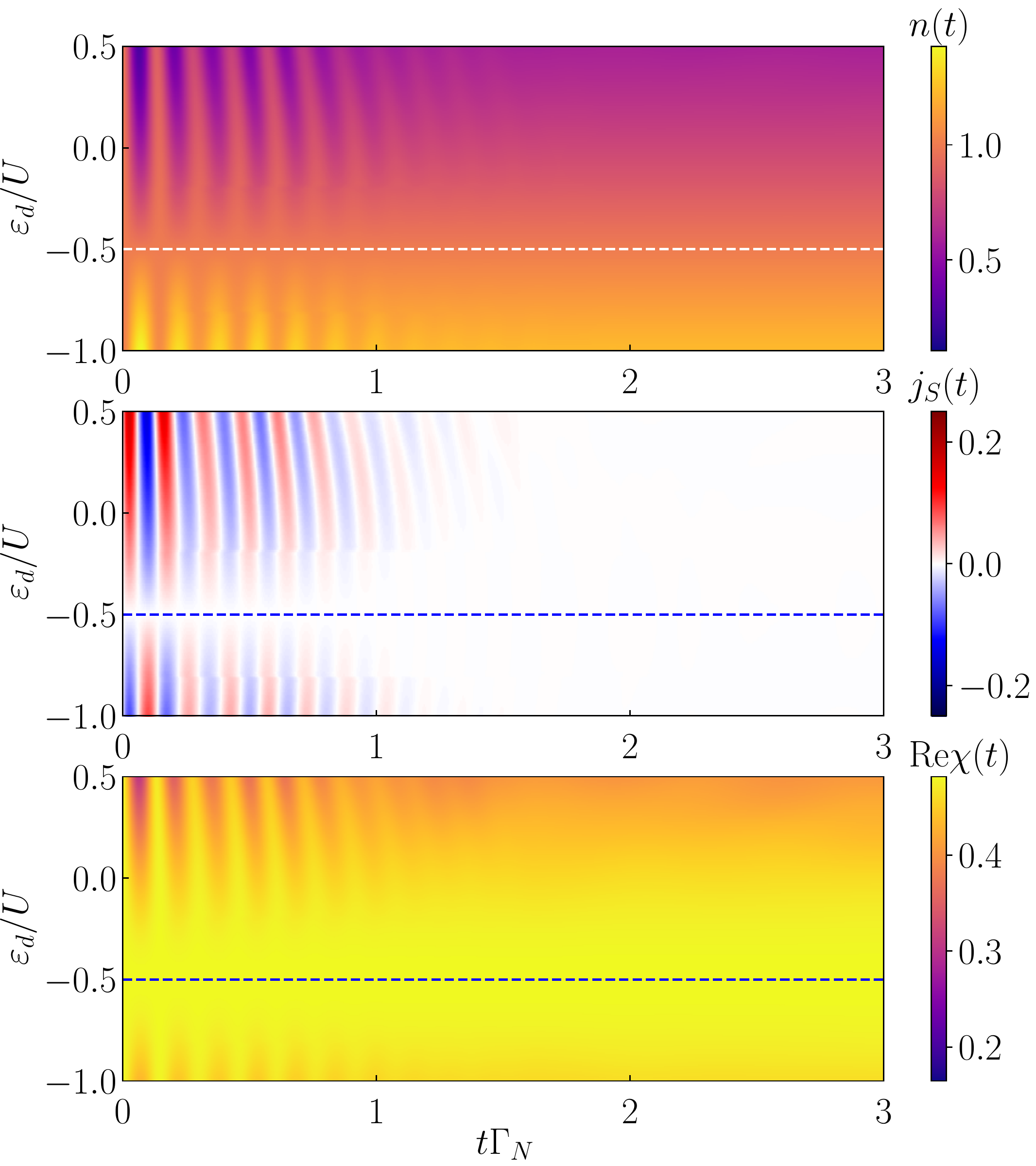}
\caption{The time dependent occupation number $n(t)$, current $j_S(t)$ and the real part of $\langle d_\downarrow(t) d_\uparrow(t) \rangle$ after QD level quench from $\varepsilon_{d}(t\leq 0)=-U/2$ to $\varepsilon_{d}(t>0)=\varepsilon_d$ as a function of time.
The coupling to superconductor is set to $\Gamma_S/U=2$
and the other parameters are the same as in Fig. \ref{quench_eps}.}
\label{tnrg_eps_quench1}
\end{figure}

Let us now inspect the second type of quantum quench, related to abrupt change of the QD energy level (\ref{abrupt_gate}). Figures~\ref{tnrg_eps_quench1} and \ref{hfb_eps_quench1} present the time-dependent observables obtained for the same parameters as in Fig.~\ref{quench_eps} by tNRG and mean field approximation, respectively. Here, we set the coupling to superconductor equal to $\Gamma_S=2U$. Initially the orbital level is tuned to the particle-hole symmetry point, $\varepsilon_d(t\leq 0)=-U/2$, marked by the horizontal dashed lines in the figures. The final value of $\varepsilon_{d}/U$ after the quench is indicated on the y-axis, correspondingly.

One can see that the evolution of physical observables to their new stationary values is realized through a sequence of quantum oscillations, analogously to the behavior displayed in Fig. \ref{quench_eps}. These oscillations show up for a wide range of final values of energy level $\varepsilon_{d}$. In this regard, the absolute difference $|\varepsilon_d(t\leq0) - \varepsilon_d(t>0)|$ has a strong influence on the amplitude of such oscillations. This is especially evident, when examining the time-dependence of all observables near the particle-hole symmetry point. However, exactly for $\varepsilon_{d}=-U/2$, the quantum oscillations are completely absent. We have previously provided physical reasoning for this phenomenon, inspecting the transient effects of uncorrelated system \cite{Taranko-2018}. The oscillations originate from the leakage of Cooper pairs onto the quantum dot and such processes are hardly possible when the initial configuration is exactly half-occupied. Away from the half-filling, however, the Cooper pairs can flow back and forth, what is manifested by the quantum oscillations in all observables. Their frequency depends on the energies $E_{A}$ of the bound states (see Fig.~\ref{idea}) reminiscent of the Rabi oscillations in two-level systems. The relaxation mechanism is contributed here by the coupling $\Gamma_{N}$ to a continuum of the metallic lead.

Abrupt change of the QD energy level has a considerable impact on the long-time limit of the occupation number. For instance, $n(t\rightarrow \infty)\approx 0.57$, for the quench to $\varepsilon_{d}/U=0.5$ and $n(t\rightarrow \infty)\approx 1.23$, for $\varepsilon_{d}/U=-1$, respectively. The occupancy oscillations are mostly pronounced right after the quench in the early time-interval $t \Gamma_N \lesssim 1$. As time elapses, they are exponentially suppressed with the relaxation rate $\tau \sim 1/\Gamma_N$. Interestingly, some intriguing effect can be observed in the time-dependent supercurrent $j_S(t)$, whose evolution is characterized by the oscillations shifted by $\pi$ upon crossing the half-filling $\varepsilon_{d}=-U/2$. The maxima perfectly coincide with minima around $\varepsilon_{d}=-U/2$, marked by the dashed lines. This effect resembles the $0-\pi$ {\it phase transition}, whose nature has been widely discussed in the literature for the stationary conditions  \cite{Novotny-2019,Meden-2019}. As already mentioned, the other current $j_{N}(t)$ is bounded with the QD occupancy $n(t)$ and $j_S(t)$ through the charge conservation law $j_{S}(t)+j_{N}(t)=e\frac{d n(t)}{dt}$.

The oscillatory behavior induced by the quench of QD energy level is least evident in the real part of the time-dependent order parameter $\chi(t)$. This quantity could be regarded as a qualitative measure of the on-dot pairing and indirectly affects the charge current $j_{N}$ (Sec.~\ref{Sec:biased}). Its magnitude is meaningful predominantly in the BCS-type ground state, as has been pointed out by the previous NRG studies \cite{Bauer-2007} under the stationary conditions.
The most significant variations of ${\rm Re}\chi(t)$ are realized in the short-time limit, when the occupation number $n(t)$ has its minima for quenches to $\varepsilon_{d}>0$. We once again recall, that the quantum dot is strongly coupled to superconductor ($\Gamma_S/U=2$), which firmly establishes the large value of ${\rm Re}\chi(t)$ in both the initial and final states. For this particular regime, the quench does not affect the long-time limit in a considerable way.

\begin{figure}
\includegraphics[width=1\columnwidth]{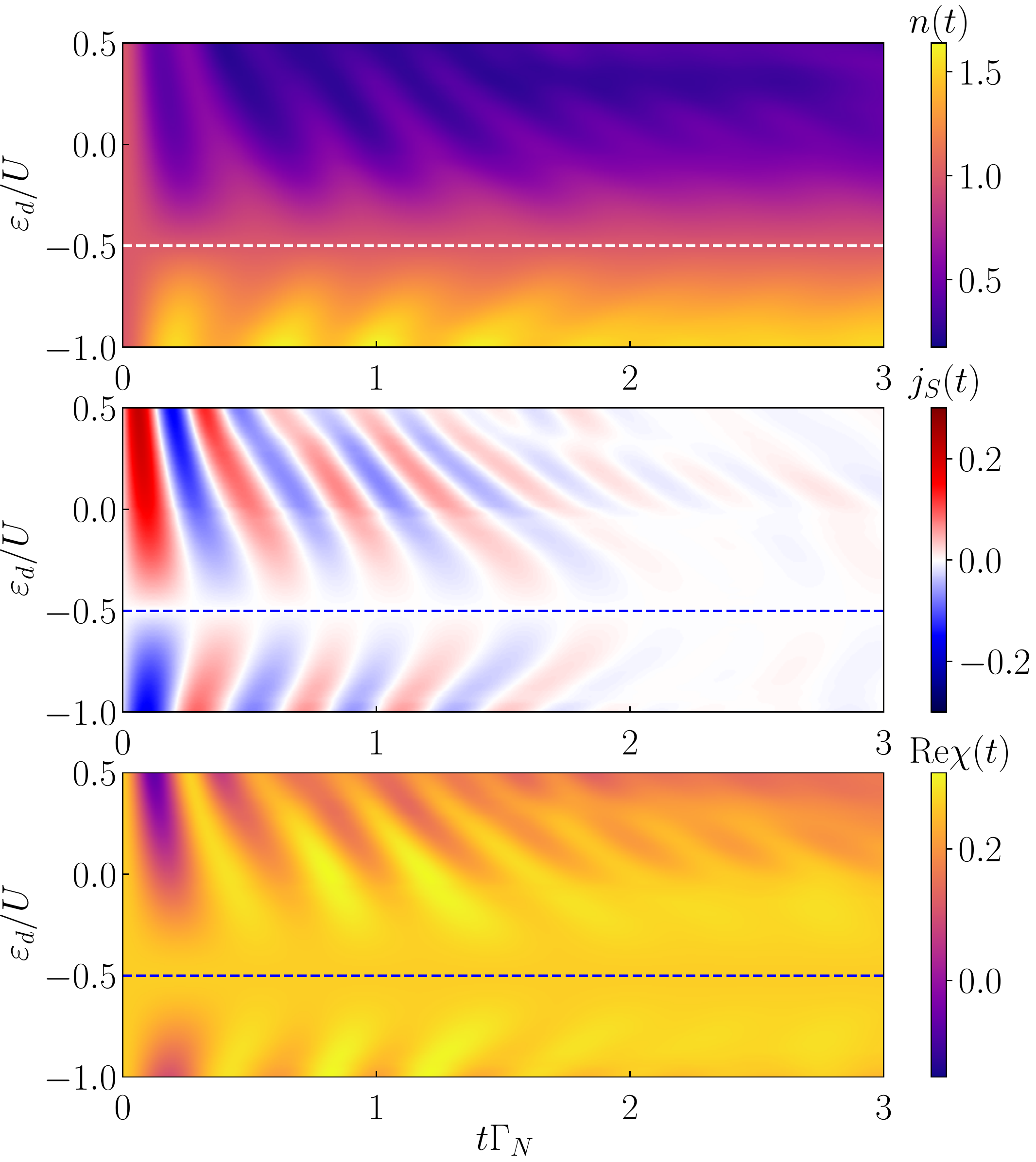}
\caption{The time dependent occupation number $n(t)$, current $j_S(t)$ and the real part of $\langle d_\downarrow(t) d_\uparrow(t) \rangle$ after QD level quench from $\varepsilon_{d}(t\leq 0)=-U/2$ to $\varepsilon_{d}(t>0)=\varepsilon_d$ as a function of time.
The coupling to superconductor is set to $\Gamma_S/U=0.5$, while
the other parameters are the same as in Fig. \ref{quench_eps}.}
\label{tnrg_eps_quench2}
\end{figure}

Further significant modifications of the oscillatory time-dependent quantities can be observed when changing the coupling to the superconductor $\Gamma_{S}$. Let us now examine typical results obtained for the system, using the parameters initially tuned to the quantum phase transition ($\Gamma_S/U=0.5$). Figure~\ref{tnrg_eps_quench2} displays the evolution obtained after the quench in the quantum dot energy level, identical to that discussed above.
All presented time-dependencies following the quantum quench maintain the oscillatory character. However, due to the reduction of the coupling strength to superconductor $\Gamma_S$, the oscillations have generally lower frequency as compared with the previous case. Additionally, the magnitude of the quench influences the frequency in the way that it is shifted toward higher values as the difference $|\varepsilon_d(t\leq0) - \varepsilon_d(t>0)|$ is increased. This behavior gives an interesting prospect for a device generating transient supercurrents with frequency controlled by appropriate switching of the gate potential $V_G$ in a step-like manner. It is important to note here that the oscillations of the imaginary part of the order parameter have the amplitude unchanged.
Smaller values of the pairing potential relax the constrains on the long-time limit for the occupation number. Here, $n(t\rightarrow \infty)\approx 0.2$ for the quench to $\varepsilon_{d}/U=0.5$ and $n(t\rightarrow \infty)\approx 1.55$ for $\varepsilon_{d}/U=-1$, which are the values spanning wider range of $n(t\rightarrow \infty)$ than in the case of the system strongly coupled to superconductor with $\Gamma_S/U=2$, cf. Fig.~\ref{tnrg_eps_quench1}. On the other hand, as expected upon lowering the pairing amplitude, the real part of the order parameter $\chi(t)$ reveals reduced values from a smaller range, both during the time evolution and after achieving the long-time limit.

\subsection{Dynamical susceptibility}

\begin{figure}
\includegraphics[width=1\columnwidth]{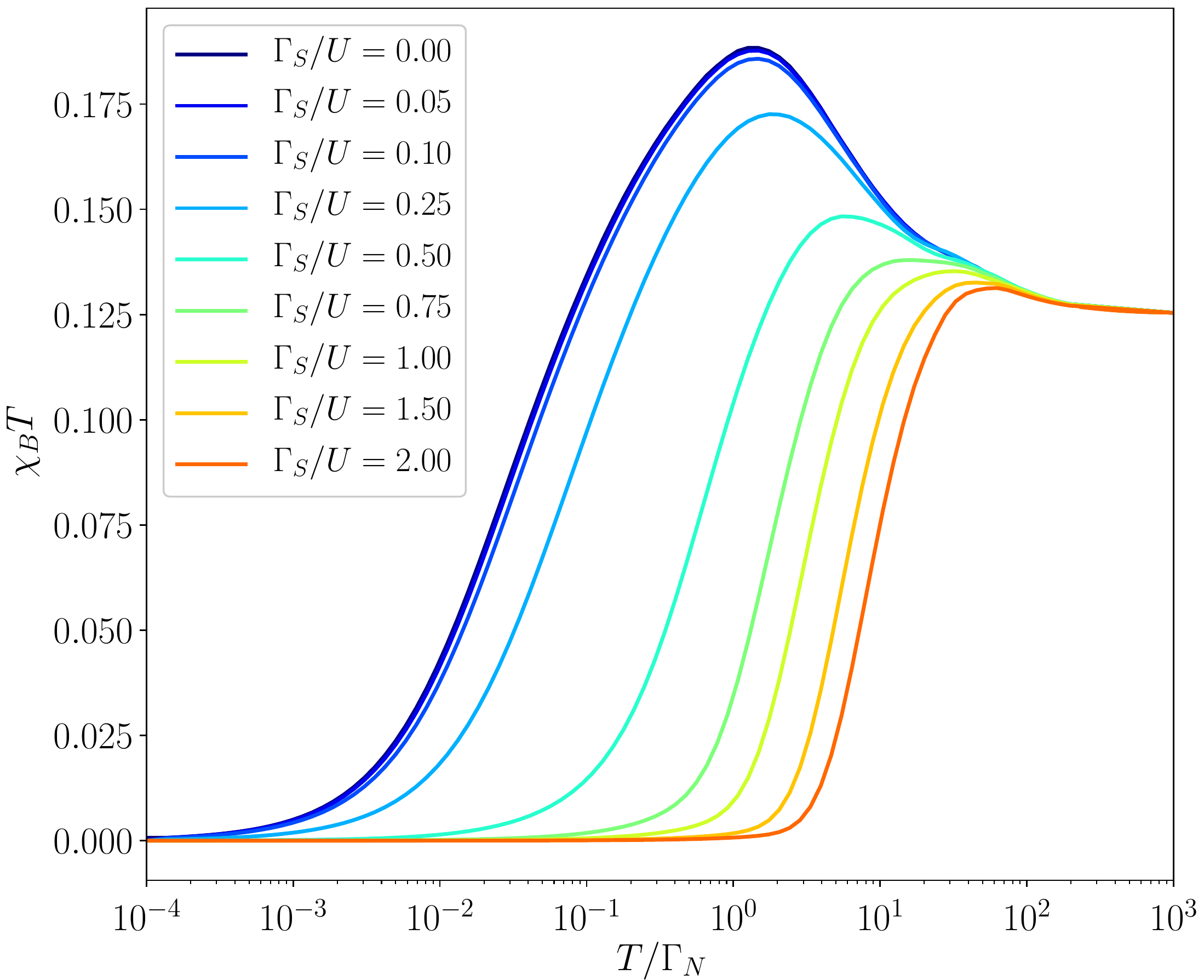}
\caption{The magnetic susceptibility as a function of temperature $T$ for several values of coupling to superconductor $\Gamma_S$, as indicated.
The other parameters are the same as in Fig. \ref{quench_gammas}.
Susceptibility is multiplied by temperature $T$.}
\label{tnrg_suscept_static}
\end{figure}

The aforementioned non-trivial dynamical behavior of the studied system can be further revealed when inspecting interplay of the superconducting correlations with the local magnetism. To get an insight into such competition, let us first examine the magnetic susceptibility defined as $\chi_B \equiv \frac{d}{dB} \langle S_z \rangle_{B=0}$ for the system in the equilibrium. Figure~\ref{tnrg_suscept_static} presents the behavior of $\chi_B$ as a function of temperature $T$ for different values of coupling to the superconducting lead $\Gamma_S$.
For the quantum dot completely decoupled from the superconductor $\Gamma_S=0$, the maximum of magnetic susceptibility is found for temperature $T\approx\Gamma_N$. It acquires reduced value of $\chi_B T \approx 0.19$ as compared with a free-spin case, where $\chi_B T = 1/4$. When the temperature decreases, the Kondo effect becomes enhanced, which results in a full screening of the quantum dot spin for $T/\Gamma_N<10^{-3}$, where $\chi_B T \rightarrow 0$. However, when the system is coupled to the superconducting lead, the temperature-dependent susceptibility is substantially modified. As the coupling $\Gamma_S$ is enhanced, the maximum of susceptibility is reduced and shifted toward higher temperatures. Moreover, the full screening of the orbital level holds at significantly higher temperatures due to the strong superconducting correlations \cite{Domanski-2016}. Finally, for high temperatures, exceeding the values of coupling and Coulomb correlations ($T>\Gamma_S,\Gamma_N, U$), all lines converge near $\chi_B T \approx 0.125$.

\begin{figure}[t]
\includegraphics[width=1\columnwidth]{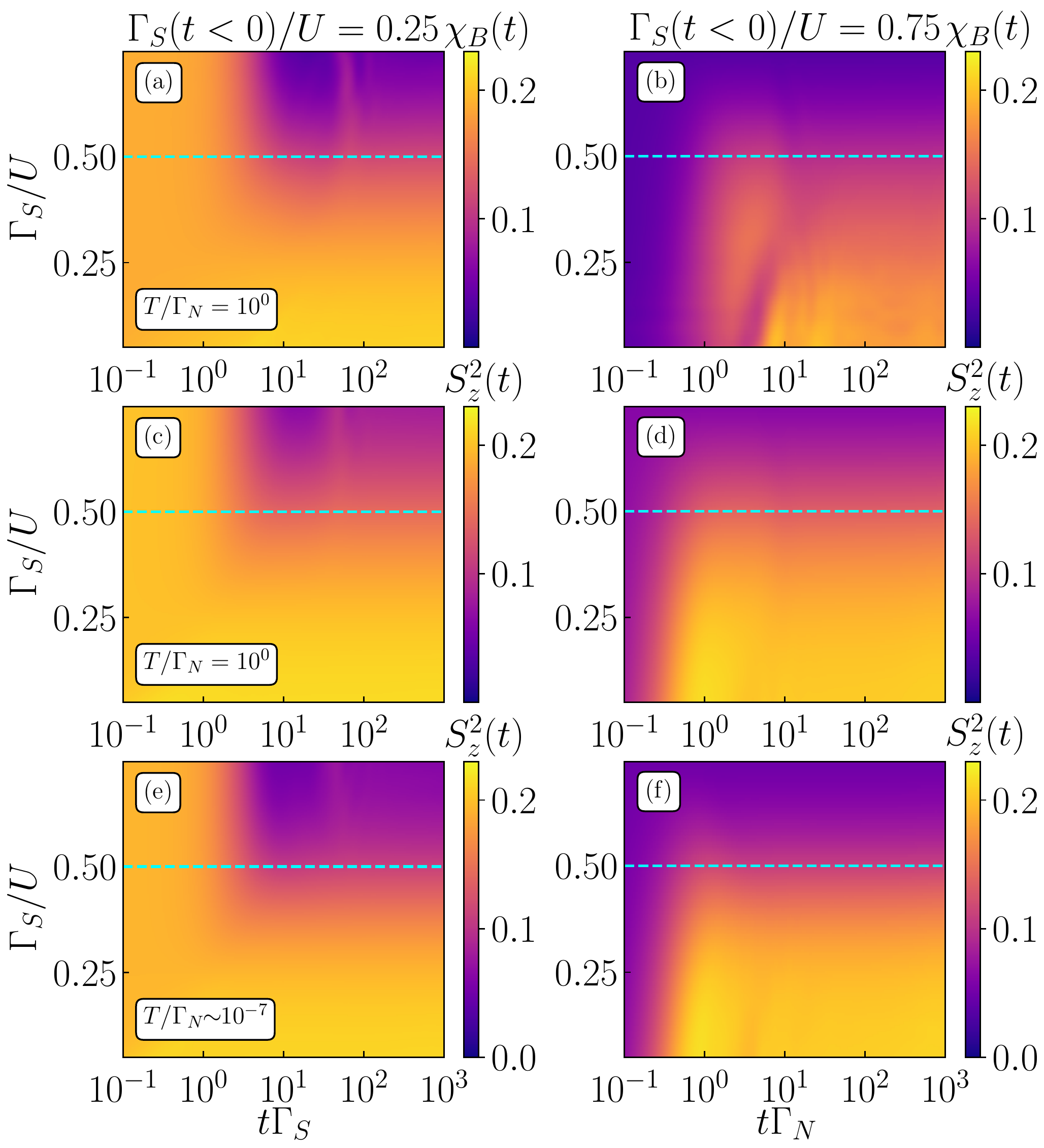}
\caption{The time dependent susceptibility $\chi_B(t)$
and square of the magnetization $S_z^2(t)$
after the quench in $\Gamma_S$ from initial value indicated at the top of each column.
Results shown in (a)-(d) are calculated for temperature $T/\Gamma_N=10^0$,
while (e) and (f) are determined for $T/\Gamma_N \sim 10^{-7}$.
Cyan dashed lines indicate the coupling strength $\Gamma_S(t)=U/2$
associated with the quantum phase transition. The other parameters are the same as in Fig. \ref{quench_gammas}.}
\label{tnrg_sd_suscept}
\end{figure}

Upon varying the coupling strength $\Gamma_S$, the most pronounced change of magnetic susceptibility occurs at temperature $T\approx\Gamma_N$. To get a better understanding of the dynamical aspects of this dependence, in Fig.~\ref{tnrg_sd_suscept} we show the time-dependent susceptibility and the squared magnetization following the quench in the coupling strength $\Gamma_S$. It is important to note, that magnetic susceptibility (being a measure of a response to external magnetic field) is a property of the system well specified at equilibrium. Here, we estimate its time evolution by calculating the magnetization in a very small but finite external magnetic field $B_z$, which allows us to approximate the time dependence of the susceptibility as $S_z(t) \approx \chi_B(t)T$.

We consider two initial values $\Gamma_S(t<0)/U=0.25$ (left column) and $\Gamma_S(t<0)/U=0.75$ (right column), associated with the earlier discussed spinful and spinless phases, respectively.  We also remind that for energy of the orbital level tuned to the particle-hole symmetry point $\varepsilon_d=-U/2$, charge and supercurrent dynamics are fully suppressed.
Let us first focus on the case when the time evolution is determined after the quench from spinful phase with initial value $\Gamma_S(t<0)/U=0.25$, see the left column in Fig.~\ref{tnrg_sd_suscept}. When the final value of the coupling strength to superconductor is chosen in a way that the system remains in the same phase, i.e. $\Gamma_S(t>0)/U<0.5$, both $\chi_B(t)$ and $S_z^2(t)$ [shown in panel (a) and (c), respectively] are monotonically evolving in a rather moderate manner to a new, slightly modified long time limit in agreement with the final thermal expectation values. This regime is contained below the cyan dashed lines indicating the crossover between the phases. However, when $\Gamma_S(t>0)/U>0.5$ (a range of $\Gamma_S$ values above the cyan line), the system undergoes a transition to a spinless phase and the time dependencies reveal a rapid drop of the magnetic properties at time $t\Gamma_S \sim 10^1$. Qualitatively, for the considered system both quantities $\chi_B(t)$ and $S_z^2(t)$ have a very similar time-dependencies and only small differences are exposed mainly due to distinct thermal expectation values for the initial and final states. Additionally, the squared magnetization evolves in a smoother manner, while the magnetic susceptibility may undergo weak oscillations at times around $t\Gamma_S \sim 10^2$ before fully relaxing to the new final state. As a reference, in Fig.~\ref{tnrg_sd_suscept}(e) we also show $S_z^2(t)$ evaluated for $T/\Gamma_N \sim 10^{-7}$, which is in good agreement with dependencies at higher temperatures. However, $\chi_B(t)$ at $T/\Gamma_N \sim 10^{-7}$, does no longer exhibit the discussed behavior due to the full suppression of magnetic susceptibility at low temperatures, as shown in Fig.~\ref{tnrg_suscept_static}.

On the other hand, when the system is initially in the spinless phase and the coupling quench is performed from $\Gamma_S(t<0)/U=0.75$ (see right column of Fig.~\ref{tnrg_sd_suscept}), the response is significantly altered when confronted with the above-discussed case. The striking difference is that here, the dynamics no longer strongly depends on the coupling $\Gamma_S$. To clearly show this effect, we plot all the time-dependent expectation values as functions of $t \Gamma_N$. For relatively small change in the coupling strength $\Gamma_S(t>0)/U>0.5$, i.e. when following the quench the system remains in the spinless phase, the quantities sustain a mild and monotonic evolution toward new thermal limit. However, when the system undergoes a phase transition to a spinful state and $\Gamma_S(t>0)/U<0.5$, see the regime below the cyan dashed line, the rise of the magnetic susceptibility and square of magnetization is considerable. The buildup of $\chi_B(t)$ is noticeable at times $t \Gamma_N \sim 10^0$, subsequently revealing similar oscillations as in the case of transition in the opposite direction. Finally, the new long time limit is achieved for time in range $10^1 \! < \! t \Gamma_N \! < \! 10^2$, depending on the size of the quench. The dynamics of $S_z^2(t)$ is again similar to the evaluated time-dependent magnetic susceptibility, but it exhibits suppressed quantum oscillations and the buildup is considerably ahead of $\chi_B(t)$. At times $t \Gamma_N \approx 10^0$, it achieves maximum, which is quickly followed by thermalization to a new thermal value obtained for times $t \Gamma_N \ll 10^1$. Finally, the low temperature behavior of $S_z^2(t)$, see Fig.~\ref{tnrg_sd_suscept}(f), allows one to predict some dynamical magnetic behavior of the system at higher temperatures and, conversely, similar to the previously discussed quench.

\section{Biased heterojunction}
\label{Sec:biased}

\begin{figure}[t]
\includegraphics[width=1\columnwidth]{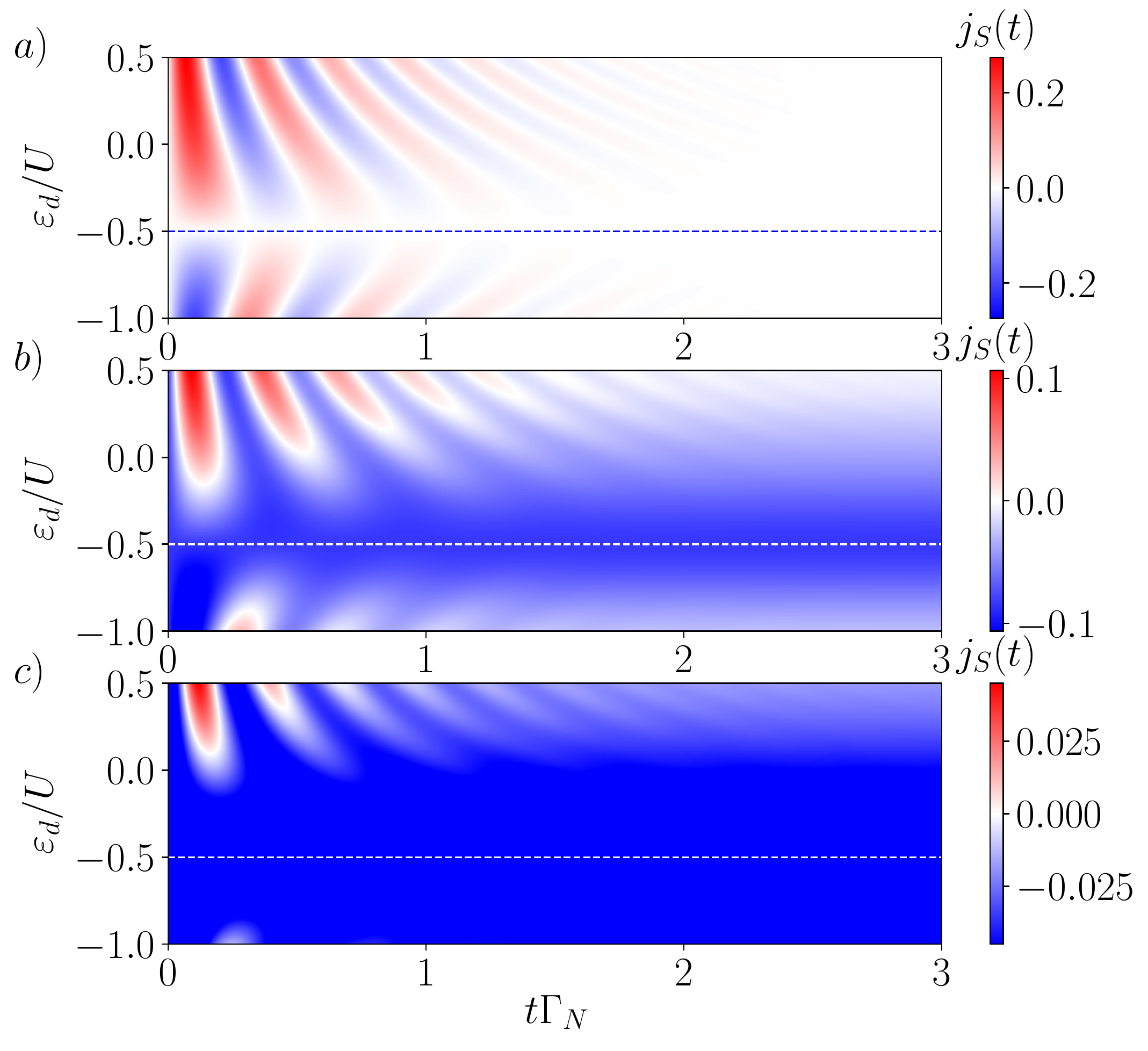}
\caption{The time-dependent current $j_{S}(t)$ obtained for the same set of parameters as in Fig.~\ref{tnrg_eps_quench2} in the presence of finite bias voltage $V$. The panels (a),(b) and (c) correspond to $eV/U=0$, $-0.5$ and $-1.0$, respectively.}
\label{hfb_nonequilibrium}
\end{figure}

Finally, we discuss the time-dependent quantities under the nonequilibrium conditions. We thus analyze the case, when the chemical potential of the normal lead $\mu_{N}$ is detuned from $\mu_{S}$  by an applied bias $eV=\mu_{N}-\mu_{S}$. For convenience we assume the superconductor to be grounded, $\mu_{S}\!=\!0$. The bias directly affects
the observables, as illustrated in Fig.~\ref{hfb_nonequilibrium}.

In what follows, we focus on the differential conductance $G_{N}(V,t)=\frac{d}{dV} I_{N}(t)$ of the charge current induced between the quantum dot and the normal lead.
The other current $j_{S}(t)$, flowing between the superconducting lead and QD, can be inferred from the charge conservation law $j_{S}(t)=\frac{dn(t)}{dt}-j_{N}(t)$.
The flow of electrons from the metallic lead to QD can be formally expressed by the following expectation value $j_{N}(t)= e \left<\frac{d}{dt} \sum_{{\bf k},\sigma}\hat{c}_{\bf k \sigma}^{\dagger}(t) \hat{c}_{\bf k \sigma}(t) \right>$. Determining the time derivative from the Heisenberg equation we can recast this current into the familiar formula
\begin{eqnarray}
j_{N}(t) = 2 e \sum_{{\bf k},\sigma} \; \mbox{\rm Im} \left\{ V_{\bf k}\left<
\hat{d}^{\dagger}_{\sigma}(t) \hat{c}_{{\bf k}\sigma}(t)\right> \right\} .
\label{current_N}
\end{eqnarray}
The second quantization operators of the metallic bath electrons are governed by $\hat{c}_{{\bf k}\sigma}(t) \!=\! \hat{c}_{{\bf k}\sigma}(0) e^{-i \xi_{\bf k} t} \!- \! i \int_{0}^{t}\!\! dt' V_{{\bf k}} e^{-i \xi_{\bf k} (t-t')} \hat{d}_{\sigma}\!(t')$. \cite{Taranko-2018}. Our main computational difficulty is related here with the time-dependent operators $\hat{d}_{\sigma}^{(\dagger)}\!(t)$. Depending on any specific type of the quantum quench these operators can be determined, applying the equation of motion procedure proposed earlier by us for investigating the dynamics of uncorrelated QD (for details see Appendix A.1 in Ref.\ \cite{Taranko-2018}).

For investigating both types of the quantum quenches we shall treat the correlations within the Hartree Fock Bogoliubov approximation (\ref{HFB}), because, as we have presented in Sec.~\ref{unbiased_junction} by comparison with tNRG, such procedure yields reliable results.

\begin{figure}[t]
\includegraphics[width=0.95\columnwidth]{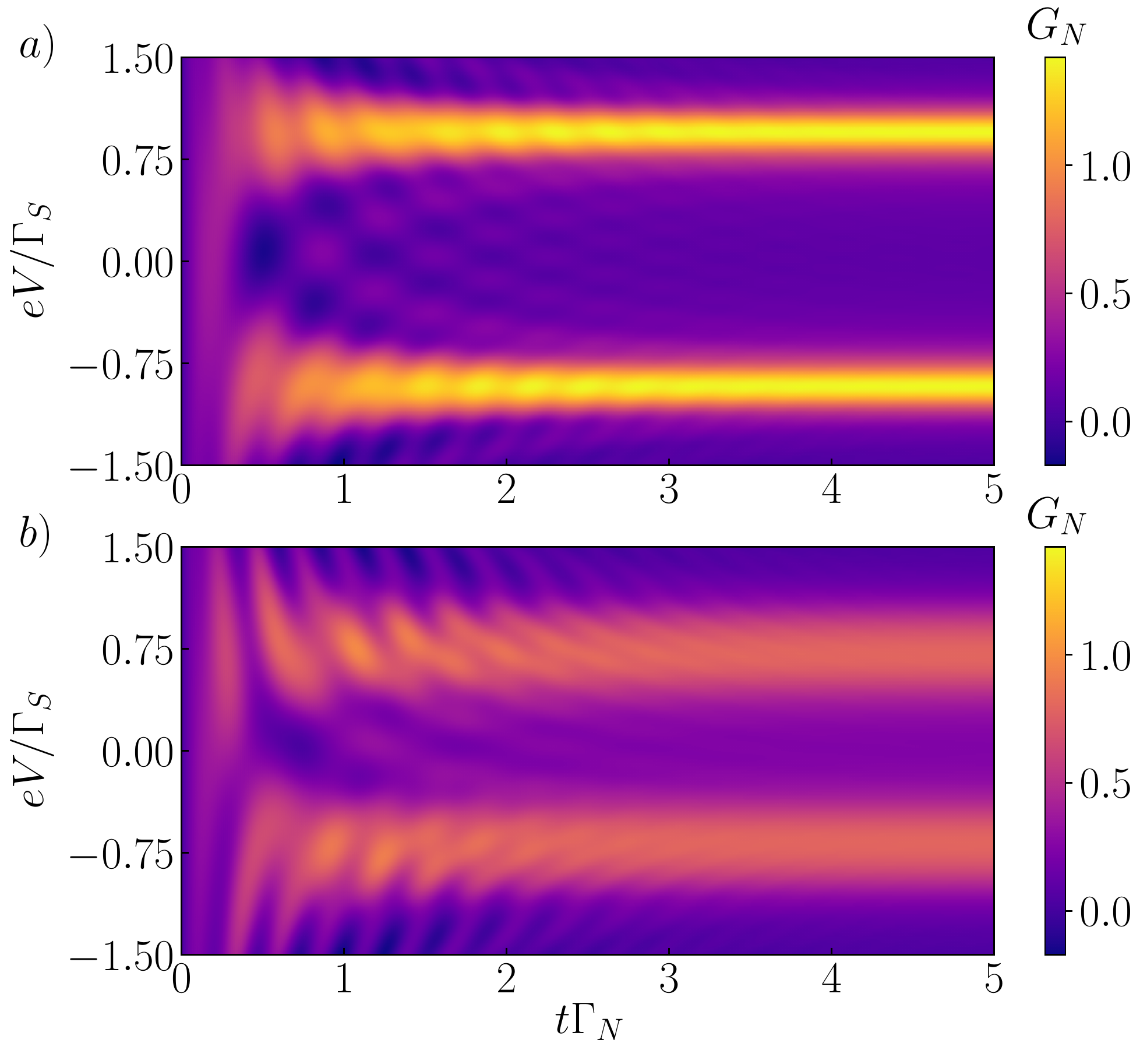}
\caption{Variation of the  differential conductance $G_{N}$ (in units of $2e^2/h$) with respect to voltage $V$ and time $t$ obtained in the mean-field approximation, imposing a sudden coupling of the QD to both external leads at $t=0^{+}$. In (a) and (b), $U=0.025$ and $U=0.1$, respectively. The other model parameters are $\Gamma_{N}=0.01$, $\Gamma_{S}=0.1$, $\varepsilon_{d}=-U/2$.}
\label{hfb_conductance_1}
\end{figure}

The steady limit value $j_{N}(\infty)$ of Eq.~(\ref{current_N}) can be independently evaluated, for instance within the Landauer formalism. Such Andreev-type spectroscopy has been widely discussed in the literature \cite{Rodero-11,Paaske-2010}. Our major interest here would be the evolution of the tunneling current $j_{N}(t)$ towards its steady-state limit, which encompasses the relaxation processes (imposed by the coupling $\Gamma_{N}$) and the quantum oscillations with frequencies sensitive to the ratio $\Gamma_{S}/U$ and dependent on the QD level $\varepsilon_{d}$.

Let us first inspect the case, when the quantum dot is  coupled at $t=0^{+}$ simultaneously to the both external leads. Under such circumstances we can observe signatures of the bound states formation manifested in the time-dependent differential conductance $G_{N}(V,t)$. Figure~\ref{hfb_conductance_1} presents these transient effects for the selected model parameters $\varepsilon_{d}$, $\Gamma_{S}$, $U$ (as indicated). These plots clearly reveal the emerging bound states around $\pm E_{A}$ of either symmetric (for $\varepsilon_{d}=-U/2$) or asymmetric spectral weights (when the quantum dot is away from its half-filling). The asymptotic features are developed at times $t \sim 1/\Gamma_{N}$ and in a meantime there occur the quantum oscillations with the period $2\pi/E_{A}$.

\begin{figure}
\includegraphics[width=0.95\columnwidth]{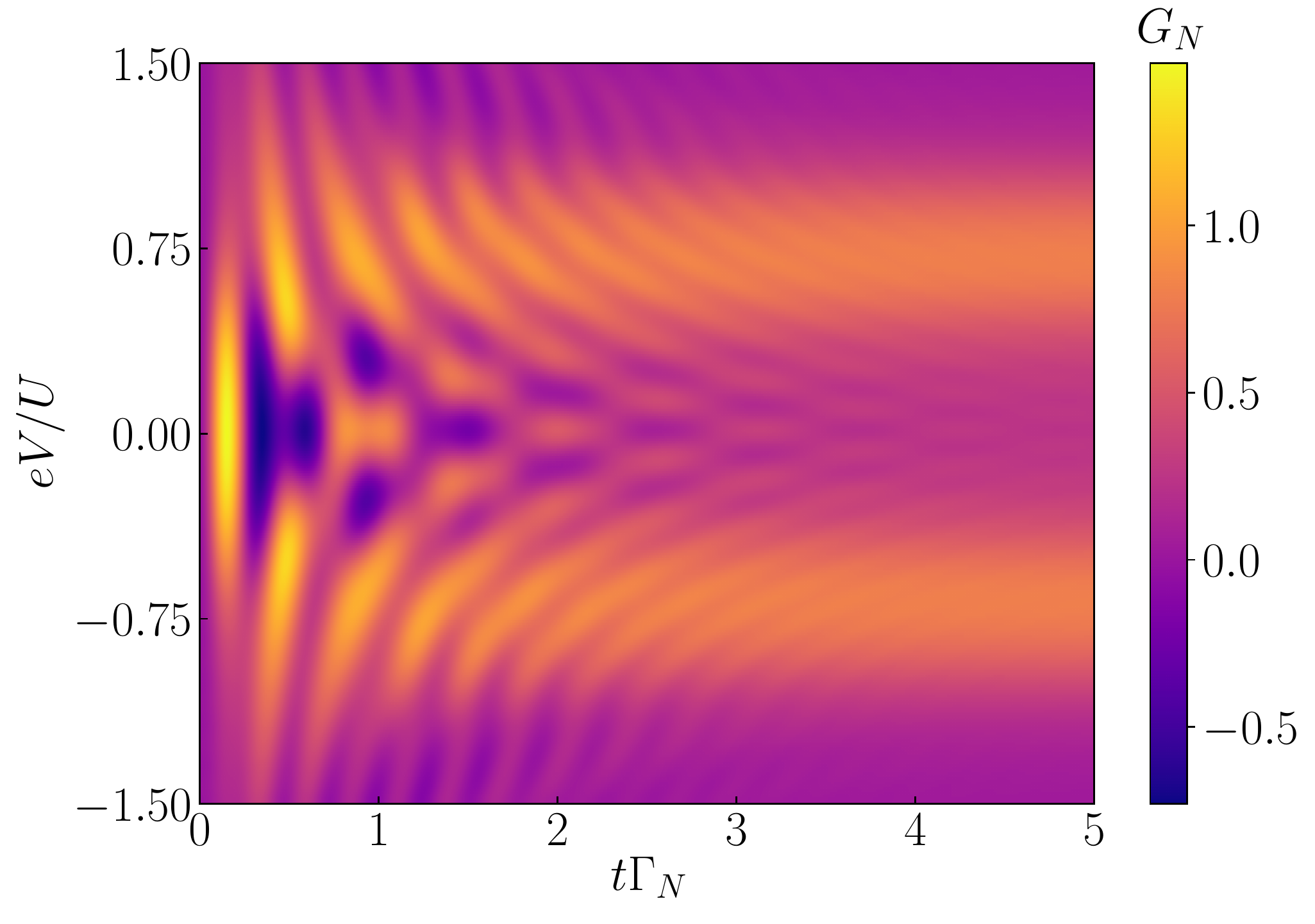}
\caption{The differential conductance $G_{N}$  as a function of voltage (vertical axis) and time (horizontal axis) obtained for the quench of hybridization $\Gamma_{S}=0\rightarrow 0.1$ at $t=0$. Calculations were done for $\Gamma=0.01$, $U=0.1$ and $\varepsilon_{d}=-U/2$.}
\label{conductance_2}
\end{figure}

Let us now turn our attention to the quantum quenches. Figure~\ref{conductance_2} displays the differential conductance obtained for the half-filled QD ($\varepsilon_{d}=-U/2$) abruptly coupled to the superconducting lead. We set the Coulomb potential $U=0.1$ and impose the quench $\Gamma_{S}(t)=U\theta(t)$. Initially the normal quantum dot is characterized by the quasiparticle peaks at energies $\varepsilon_{d}$ and $\varepsilon_{d}+U$, which for the half-filled QD occur at $\pm U/2$. The superconducting proximity effect drives the quantum dot to the new quasiparticle states at energies $\pm E_{A}$ (their values in the limit of $\Gamma_{N}=0$ are $E_{A}\sim \sqrt{(\varepsilon_{d}+U/2)^{2}+\Gamma_{S}^{2}}$). We notice, that emergence of such new quasiparictles resembles transient phenomena presented in Fig.~\ref{hfb_conductance_1}. This behavior is rather not surprising, considering that the coupling $\Gamma_{N}$ is much weaker compared to $\Gamma_{S}$ and $U$.

\begin{figure}
\includegraphics[width=0.95\columnwidth]{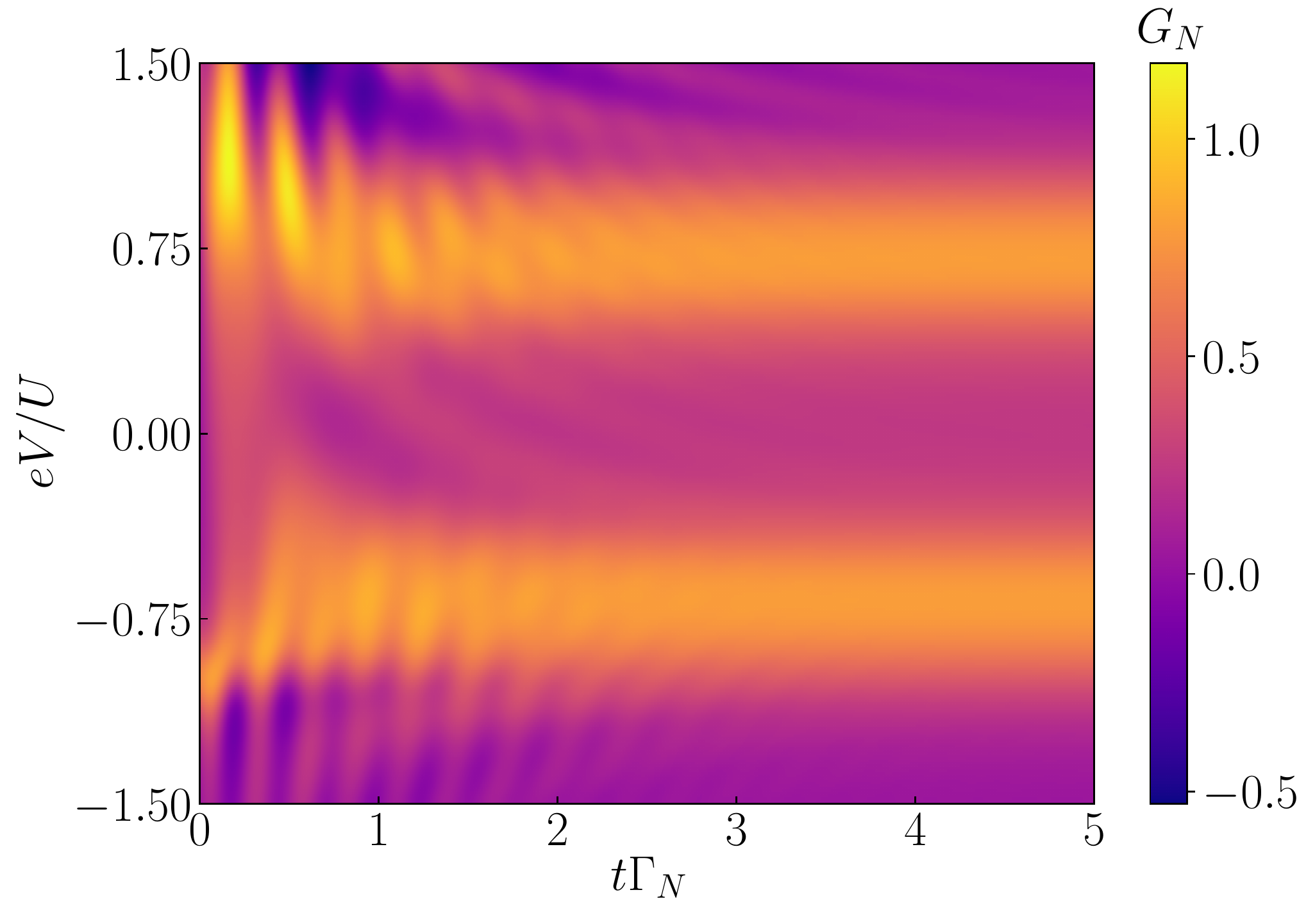}
\caption{The differential conductance $G_{N}$  obtained for $\Gamma_{N}=0.01$ and $\Gamma_{S}/U=1$, imposing a sudden change of the QD energy level  $\varepsilon_{d}=U/2\rightarrow -U/2$ at $t=0$.}
\label{conductance_4}
\end{figure}

Figure~\ref{conductance_4} shows the differential conductance obtained for the quench of the energy level, from its initial value $\varepsilon_{d}(t\leq 0)=-U/2$ to  $\varepsilon_{d}(t > 0)=U/2$. We assume $\Gamma_{S}=U$, therefore both at initial and final stages the quantum dot would be safely in the BCS-type configuration. Sudden change of the energy level is here responsible for modifying the energies $\pm E_{A}$ of subgap quasiparticles and gradual development of their asymmetric spectral weights.

\begin{figure}
\includegraphics[width=0.95\columnwidth]{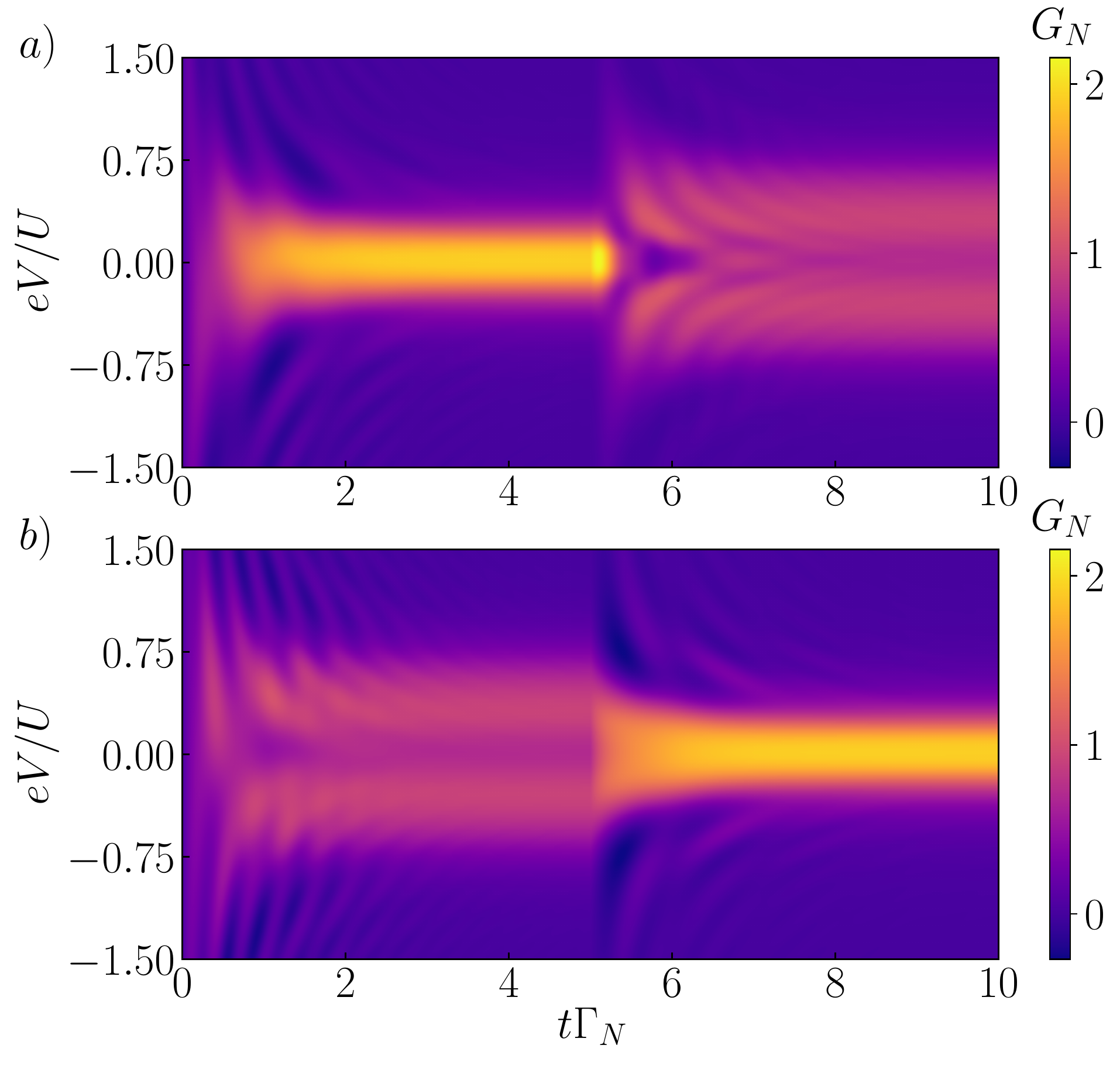}
\caption{The differential conductance $G_{N}$ (in units of $2e^2/h$) obtained across the doublet-singlet transition due to the quench of $\Gamma_{S}$: from $\Gamma_{S}=0.03$ up to $\Gamma_{S}=0.07$ (upper panel) and from $\Gamma_{S}=0.07$ down to $\Gamma_{S}=0.03$ (bottom panel). The quench was imposed at $t=5/\Gamma_{N}$ using the model parameters $\Gamma_{N}=0.01$, $U=0.1$ and $\varepsilon_{d}=-U/2$.}
\label{hfb_conductance_3}
\end{figure}

Finally, we consider the evolution of the quasiparticle spectra, in which one could observe transitions between the singlet and doublet configurations. Such situation can be achieved in two steps, as displayed in Fig.~\ref{hfb_conductance_3}. Initially, at $t=0^{+}$, the half-filled QD quantum dot is coupled to both electrodes, with $\Gamma_{S}>U/2$ (upper panel) or weakly $\Gamma_{S}<U/2$ (bottom panel). In the time interval $t \in \left( 0, 5/\Gamma_{N} \right>$ we analyze the transient effects.  Next, at $t=5/\Gamma_{N}$, we abruptly reverse  these couplings $\Gamma_{S}$. This quench triggers transitions from the doublet-to-singlet (in the upper panel) and from the singlet-to-doublet (in the bottom panel), respectively. We notice, that postquench behaviour is not completely identical for both cases but the quasiparticle features in the upper/bottom panel right before the quench are pretty similar to the asymptotic ones in the bottom/upper panels.

\section{Summary}
\label{Sec:conclusions}

We have studied the dynamical properties of the correlated quantum dot sandwiched between the metallic and  superconducting leads, considering the quantum quenches driven by (a) sudden change of the energy level and (b) abrupt variation of the coupling of the quantum dot to the superconductor. We have treated the correlations within the non-perturbative time-dependent numerical renormalization group scheme and compared such results to the Hartree Fock Bogoliubov mean field approach.
For both types of quenches, we observe that the time-dependent observables (such as quantum dot charge, complex order parameter, and local currents) gradually evolve to their stationary limit values through a series of damped quantum oscillations. Frequencies of these oscillations coincide with the energies of the in-gap quasiparticles, whereas the rate of relaxation processes depends on the dot coupling $\Gamma_{N}$ to a continuous spectrum of the metallic reservoir.

We have inspected in more detail the specific realizations of quenches, which enable a changeover of the quantum dot ground states between the singlet/doublet (spinless/spinful) configurations. Traversing from the BCS-type to the doublet configuration (and {\it vice versa}) we have noticed $\pi$-shift of the charge current flowing from the superconductor to the quantum dot $j_{S}(t)$ observable at arbitrary time $t$. It can be regarded as the time-dependent signature of the, so called, $0-\pi$ transition  reported previously under the stationary conditions for the correlated quantum dot embedded in the Josephson-type junctions \cite{Rodero-11,Zonda-2015,Meden-2019}.

We have also found qualitative changes showing up in the magnetic properties upon approaching the quantum phase transition (induced either by the quench of the energy level $\varepsilon_{d}$ or the coupling $\Gamma_{S}$). The time-dependent magnetic susceptibility and the squared quantum dot spin clearly reveal a competition between the on-dot paring and the Coulomb repulsion. Dynamical signatures of such competition are manifested also in the time-dependent order parameter.

Since practical verification of the mentioned dynamical properties could be obtained from measurements of the tunneling currents, we have investigated the time-dependent differential conductance. In particular, we have focused on the charge flow induced between the metallic lead and the dot in presence of the bias potential. We have found, that its voltage characteristics clearly reveal all the necessary details of the time-dependent subgap quasiparticles. The quantum quenches could thus be used for inspecting the energies and life-times of such in-gap quasiparticles from a dynamical perspective.

\begin{acknowledgments}
This work is supported by the National Science Centre (Poland) under the grants 2017/27/B/ST3/00621 (KW, IW)
2017/27/B/ST3/01911 (BB, RT), and 2018/29/B/ST3/00937 (TD).
\end{acknowledgments}

\appendix

\section{Mean field results}
\label{HFB_results}

In this appendix we present the time-dependent observables obtained within the mean field
approximation for the unbiased heterostructure, using the same set of model parameters
as in tNRG calculations which have been discussed in Sec.~\ref{unbiased_junction}.
We could observe very good agreement, both qualitatively and quantitatively.

\begin{figure}
\includegraphics[width=1\columnwidth]{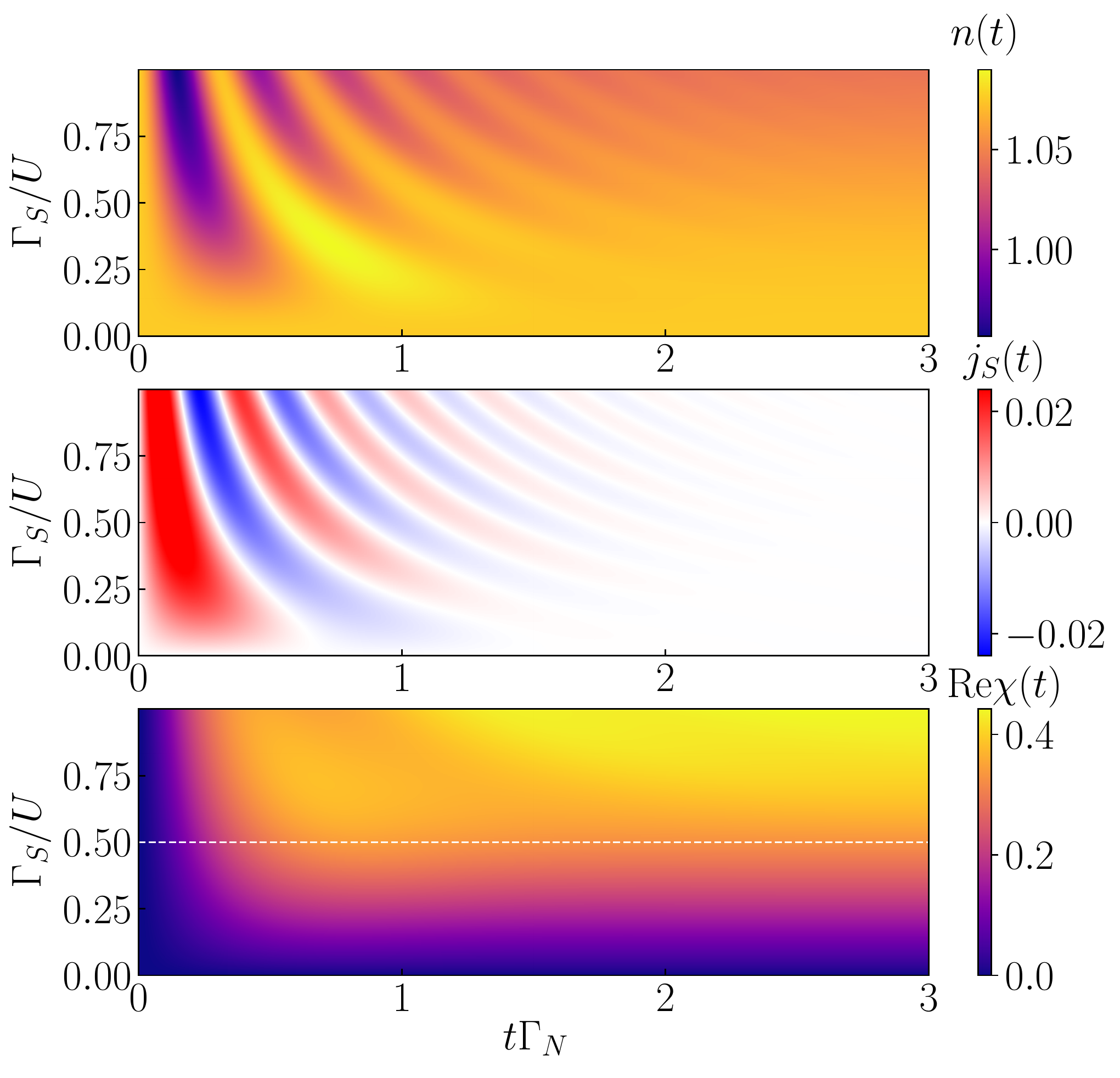}
\caption{Same as in Fig.~\ref{tnrg_gs_quench3} obtained by HFB approximation.}
\label{hfb_gs_quench3}
\end{figure}

\begin{figure}
\includegraphics[width=1\columnwidth]{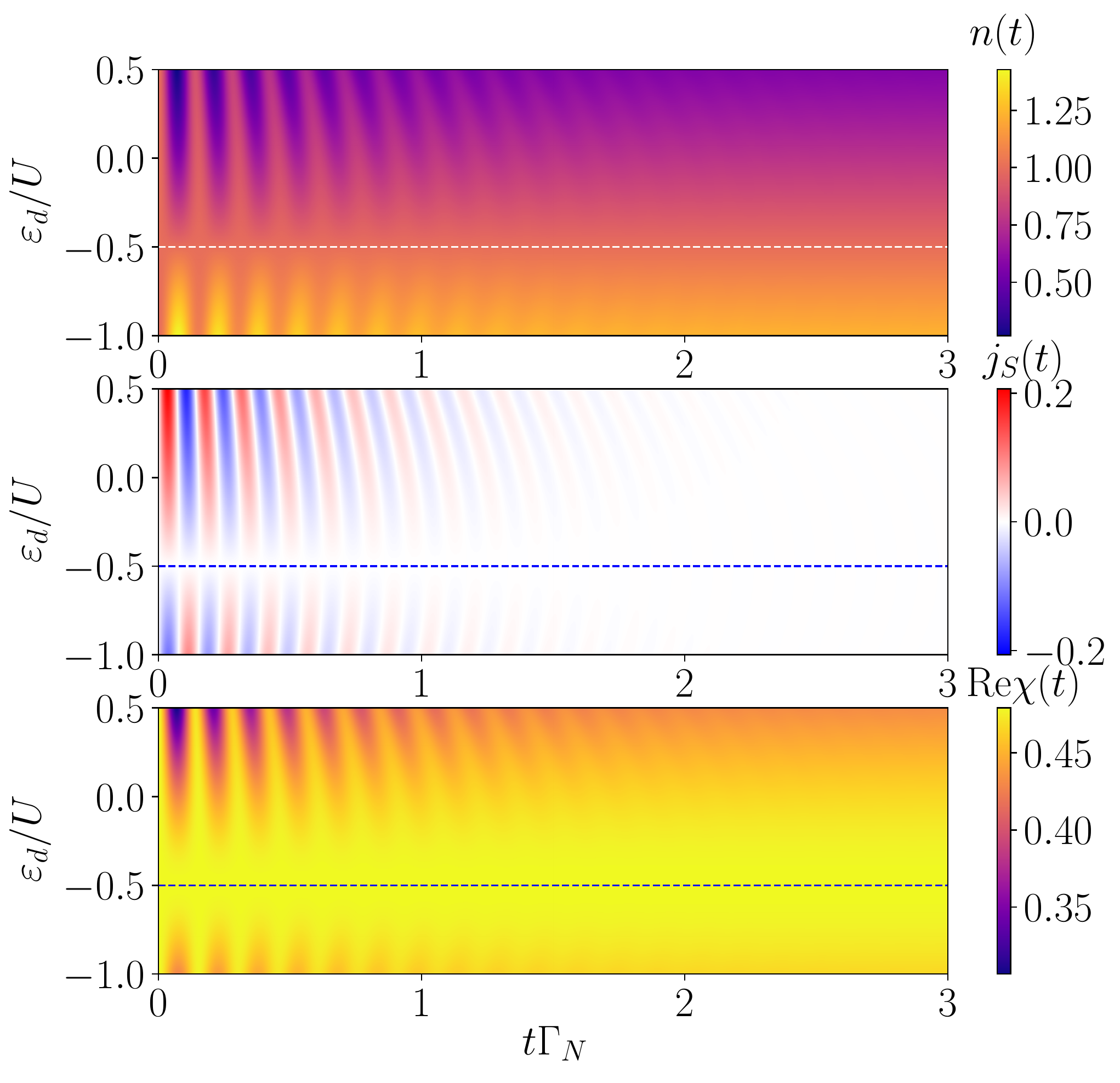}
\caption{Same as in Fig.~\ref{tnrg_eps_quench1} obtained by HFB approximation.}
\label{hfb_eps_quench1}
\end{figure}

%

\end{document}